\documentclass[twocolumn,showpacs,prb,amsmath,amssymb,floatfix]{revtex4}
\usepackage{graphicx,bm,color,subfigure}
\usepackage[colorlinks=true,linkcolor=blue,citecolor=blue,urlcolor=blue]{hyperref}
\DeclareMathOperator{\Tr}{Tr}

\begin{document}

\title{Flux-vector model of spin noise in superconducting circuits: Electron versus nuclear spins and role of phase transition}
\author{S. LaForest}
\author{Rog\'{e}rio de Sousa} \email[]{rdesousa@uvic.ca} 
\affiliation{Department of Physics and Astronomy, University of Victoria, Victoria, British Columbia V8W 2Y2, Canada}
\date{\today}

\begin{abstract}
  Superconducting Quantum Interference Devices (SQUIDs) and other
  superconducting circuits are limited by intrinsic flux noise with
  spectral density $1/f^{\alpha}$ with $\alpha<1$ whose origin is
  believed to be due to spin impurities.  Here we present a theory of
  flux noise that takes into account the vectorial nature of the
  coupling of spins to superconducting wires.  We present explicit
  numerical calculations of the flux noise power (spectral density
  integrated over all frequencies) for electron impurities and lattice
  nuclear spins under several different assumptions. The noise power
  is shown to be dominated by surface electron spins near the wire
  edges, with bulk lattice nuclear spins contributing $\sim 5$\% of the noise power in
  aluminum and niobium wires.  We consider the role of electron spin
  phase transitions, showing that the spin-spin correlation length
  (describing e.g.  the average size of ferromagnetic spin clusters)
  greatly impacts the scaling of flux noise with wire geometry.
  Remarkably, flux noise power is exactly equal to zero when the spins are polarized along the flux vector direction, forming what we call a poloidal state. Flux noise is non-zero for other spin textures, but gets reduced in the presence of correlated ferromagnetic fluctuations between the top and bottom wire surfaces, where the flux vectors are antiparallel. This demonstrates that engineering spin textures and/or inter-surface correlation provides a method to reduce flux noise in superconducting devices.
\end{abstract}

\pacs{85.25.Dq, 
05.40.-a, 
85.25.Am} 

\maketitle

\section{Introduction\label{sec:intro}}

Superconducting Quantum Interference Devices (SQUIDs) are among the
most sensitive and useful magnetometers.\cite{clarke11} They are able
to detect magnetic fields as low as $10^{-17}$~Tesla,\cite{everitt11}
and are currently used in a wide variety of applications. Examples
include outer space tests of general relativity,
detection of short-circuit faults in microchips, as well as several
applications in medicine, such as measuring regions of brain activity
in magneto-encephalography.  However, this high degree of sensitivity
also causes the SQUIDs to be sensitive to magnetic fluctuations
intrinsic to its wires and interfaces.

The most sensitive SQUIDs show excess flux noise (in addition to Johnson-Nyquist white noise) of the order of
$1\;\mu\Phi_0/\sqrt{\textrm{Hz}}$ at a frequency of $1$~Hz
[$\Phi_0=h/(2|e|)$ is the flux quantum].  This value has not changed in order of magnitude
since the first measurements of flux noise in the 1980's;
\cite{koch83,wellstood87} only minor improvements are observed in
modern devices.\cite{yoshihara06,bialczak07,lanting09,anton13,yoshihara10,yan12} While
flux noise is sufficiently low for several applications, it is
still considered a barrier for the use of the SQUID as a quantum
bit (the flux qubit) in a superconductor-based quantum computer architecture.\cite{clarke08}
Flux noise induces dephasing and relaxation of flux qubits,
limiting their coherence times to less than
$20$~$\mu$s.\cite{stern14} Flux noise also degrades performance of other superconducting (SC) qubits such as the transmon\cite{slichter12} and phase qubit.\cite{bialczak07}
The effort to reduce flux noise and
increase qubit coherence times has been a major source of motivation
for research in improving SC devices.

\begin{figure}
\includegraphics[width=0.49\textwidth]{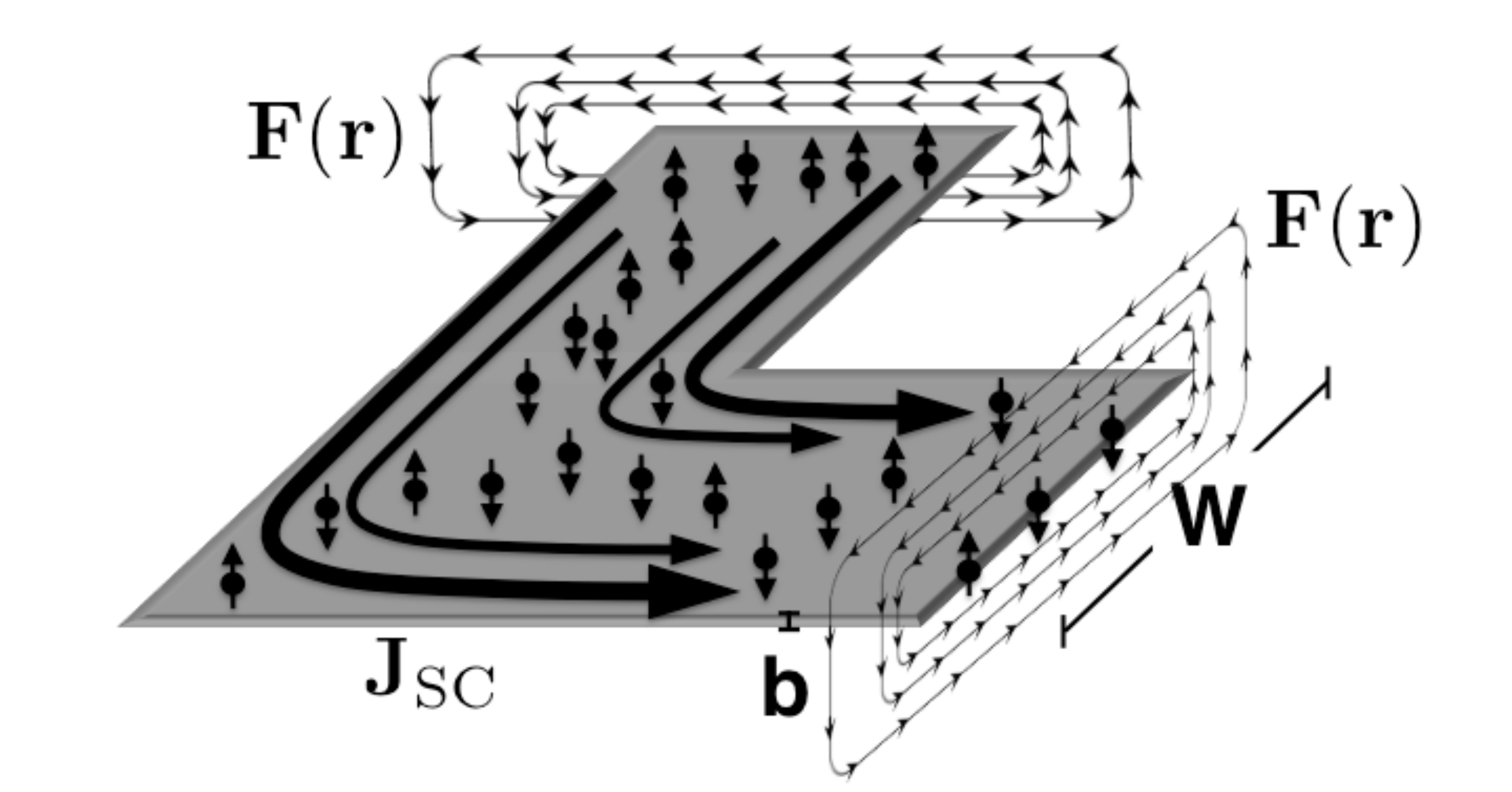}
\caption{Section of a superconducting wire with a random distribution
  of spin impurities.  The wire is made of superconducting thin-film
  of thickness $b$ and lateral width $W$, with $b\ll W$.  The flux
  $\Phi_i$ produced by a spin $\bm{s}_i$ located at position
  $\bm{R}_i$ is given by $\Phi_i=-\bm{F}(\bm{R}_i)\cdot \bm{s}_i$, with
  flux vector $\bm{F}(\bm{R}_i)$ parallel to the magnetic
  field produced by the supercurrent at $\bm{R}_i$.}
\label{FigF}
\end{figure}

\begin{figure}
\includegraphics[width=3.4in]{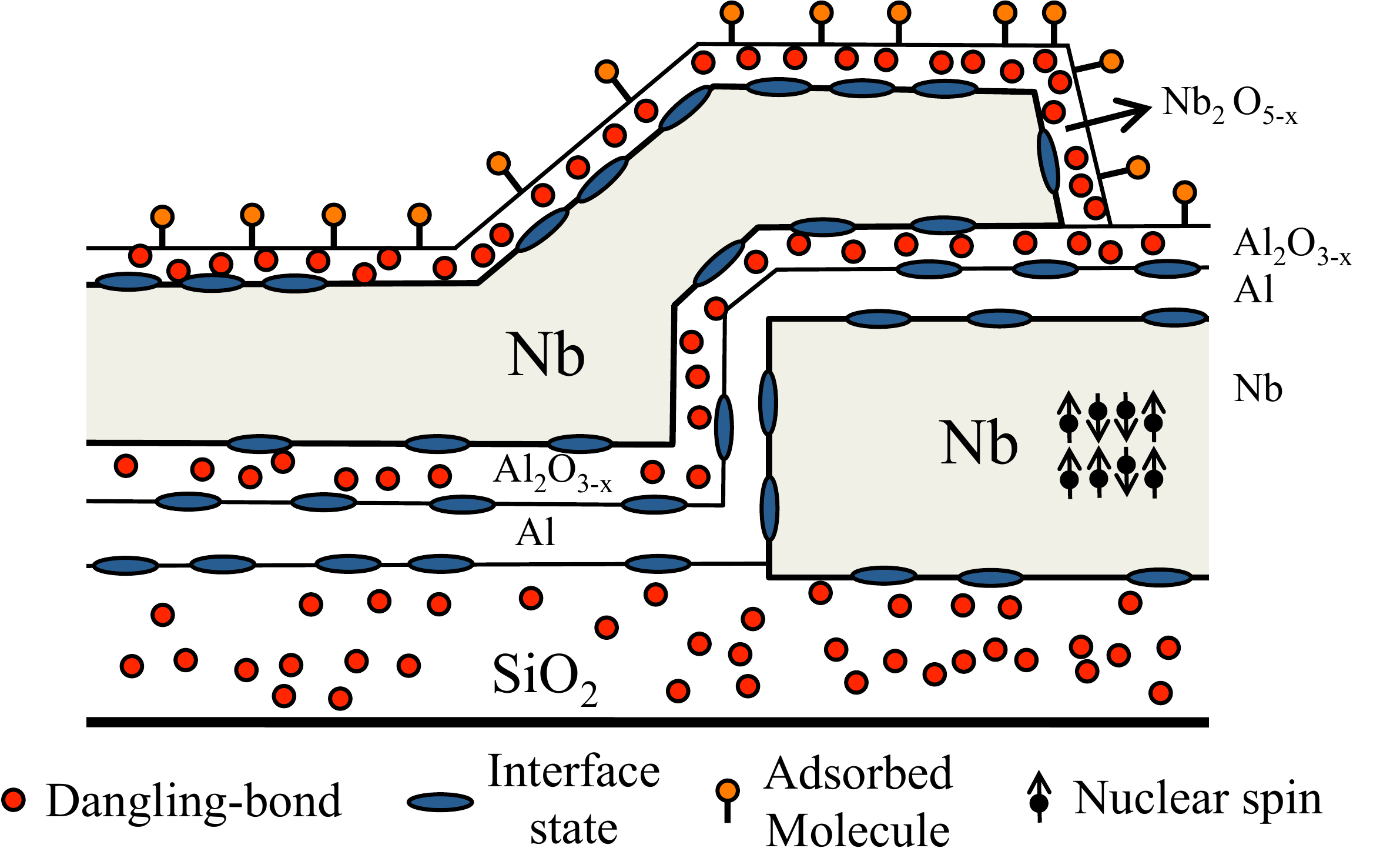}
\caption{(color online). Longitudinal cross-section of a typical
  Josephson junction\cite{lanting09,lanting14} and expected location
  of the spin species causing flux noise.  The superconducting wire is
  made of niobium, the insulator of partially oxidized aluminum
  (likely amorphous), and the substrate of silicon-dioxide.
  Candidates for the spin species include electron and nuclear spins.
  Amorphous interfaces are well known sources of electron spin centers
  such as dangling-bonds\cite{desousa07} and interface
  states.\cite{choi09} We also depict molecules adsorbed to the
  surface, that were shown to lead to electron traps.\cite{lee14} All
  materials have non-zero lattice nuclear spin that contribute to flux
  noise even in the absence of defects (e.g. Nb has nuclear spin
  $S=9/2$).}
\label{Figcrosssection}
\end{figure}

The flux threading a SQUID at low temperatures was shown to follow the
Curie susceptibility law (magnetic susceptibility $\propto 1/T$),\cite{sendelbach08}
supporting the idea that the origin of flux noise was due to the
fluctuation of local magnetic moments (presumably spin impurities) 
near the superconducting wire (Fig.~\ref{FigF}).\cite{koch07,desousa07,faoro08,choi09,chen10}
The identity of these local moments remains unknown; some
of the possibilities are illustrated in Fig.~\ref{Figcrosssection},
which depicts the longitudinal cross-section of a typical Josephson junction.
Candidates for the local moments include a variety of spin species: 
Dangling-bonds,\cite{desousa07} interface states,\cite{choi09} adsorbed molecules,\cite{lee14}
and nuclear spins of all atoms forming the materials.\cite{sleator85,dube01,rose01,wu12}

Recent theory and experiment \cite{lanting14} provided evidence that
low temperature/low frequency SQUID flux noise can be explained by a spin diffusion
model, with spin diffusion constant measured to be of the order of
$10\;\mu$m$^{2}/$s. Interestingly, this value is right in the range of
spin diffusion constants measured for lattice nuclear spins due to
their mutual dipolar interaction.\cite{zhang98} 
The typical metals that form superconducting wires (e.g.
niobium, aluminum) all have non-zero lattice nuclear spin, that are expected to contribute to 
intrinsic flux noise up to frequencies of the order of $10^{4}$~Hz (the value of dipolar coupling between nearest-neighbor nuclei in the crystal lattice).\cite{desousa09}
Nuclear spin noise should be present even in  hypothetically ``perfect'' devices that contain no electron spin centers. 

Recent experiments\cite{slichter12,bylander11} claimed the observation of flux noise at frequencies several orders of magnitude above the nuclear spin cut-off frequency ($10^{4}$~Hz). Therefore, nuclear spins alone can not explain the origin of flux noise in SC devices.

Here we describe a theory for the excess flux detected by
SC circuits in the presence of localized magnetic moments, and make explicit numerical predictions for the flux noise power (noise
spectral density integrated over all frequencies) originating from electron and nuclear spins
uniformly distributed in the surface and bulk of the SC wires. 
Our main assumption is that the superconducting condensate affects the value of the spin's magnetic moment only through local screening described by the London equations. 
We consider the impact of the formation of spin clusters (spatial
spin-spin correlation), which is typical of electron spin systems
close to a phase transition.\cite{atalaya14,de14}  This allows the assessment of the
relative contributions of surface impurity electrons and bulk lattice
nuclear spins in a variety of regimes. We present simple analytic expressions 
that allow direct comparisons to flux noise models and experiments.

As we shall demonstrate, flux noise power depends crucially on the
vectorial nature of the coupling between SC circuit to spins.  This
vectorial nature can be mapped out by measuring flux noise as a
function of the direction and magnitude of an external magnetic field
applied along the plane of the wires.

Previous calculations of flux noise due to localized spin\cite{koch07}
were based on modelling the spin as a square loop of side $0.1\;\mu$m,
the minimum feature size allowed by the finite element software
FastHenry.\cite{smithhisler96} More recently, extensive numerical
studies\cite{anton13b, lanting14} showed that these calculations
greatly underestimated the value of the flux produced by spins located
at the wire surfaces, because they did not take account of the
singular nature of the spin's dipole field. Indeed, analytic
expressions for the flux noise power that take into account the spin's
singular behaviour are still absent from the literature.  Below we
obtain these expressions and argue that they can be applied to
arbitrary circuits made of thin-film wires. We shall show below that
our expressions lead to much larger values than previous
ones,\cite{bialczak07} but are in close agreement with the most recent
numerical studies.\cite{anton13b}

The article is organized as follows: Section~\ref{sec:fluxvector}
introduces our flux-vector model, and develops a general theory on how
flux noise depends on flux vector and the formation of spin clusters
near a phase transition.  Section~\ref{sec:numermeth} describes our
numerical dipole method to calculate the flux vector explicitly.  
Section~\ref{sec:numerres} describes our numerical results and
compares them to FastHenry calculations.
Section~\ref{sec:noisepowersurface} describes numerical results for
the noise power due to electron spins distributed in the surface of
the wires, with and without spin clustering.
Section~\ref{sec:noisepowerbulk} describes our results for spins
distributed in the bulk of the wire, with explicit calculations for
lattice nuclear spins in aluminum and niobium.
Section~\ref{sec:conclusions} discusses the implications of our
results for reducing flux noise and presents our conclusions.

\section{Flux-vector model\label{sec:fluxvector}}

\subsection{General case}

Consider an ensemble of localized spins labelled by $i=1,\ldots, N$.
Each spin is located at position $\bm{R}_i$, and is described by the spin-$S$ operator $\bm{s}_i$. This can describe electron or nuclear spins, e.g. single electron impurity centers ($S=1/2$), many-electron transition metal centers ($S\geq 1/2$), or nuclear spins of lattice atoms such as 
aluminum ($S=5/2$) or 
niobium ($S=9/2$). We introduce the notion of the flux vector
$\bm{F}_i\equiv\bm{F}(\bm{R}_i)$, whose components $F_{i\alpha}$ describe the
value of the flux for a spin pointing along direction $\alpha=x,y,z$. The total
flux that the spin produces on the wires forming a device is written as
\begin{equation}
\Phi=-\sum_i \bm{F}_i\cdot \bm{s}_i.
\label{phitot}
\end{equation}
Note that the $\bm{F}_i$ are real vectors with dimensions of flux ($\bm{s}_i$ is assumed dimensionless), and $\Phi$ is a scalar quantum operator describing flux. 
The sign in Eq.~(\ref{phitot}) ensures $\bm{F}_i\parallel \bm{B}(\bm{R}_i)$, the 
magnetic field produced by the SC's current density at the spin's location $\bm{R}_i$ (See Section~\ref{sec:numermeth} below).
Such a flux is directly measured in a SQUID, but more generally will couple to any superconducting circuit by producing a voltage $V=-d\Phi/dt$.
The problem
of flux noise in superconducting circuits is to compute the thermal equilibrium noise
spectral density:
\begin{equation}
\tilde{S}_{\Phi}(f)=\int_{-\infty}^{\infty}dt\;\textrm{e}^{2\pi i f t}\left\langle\delta\Phi(t)\delta\Phi(0)\right\rangle,
\label{stildedef}
\end{equation}
where $\delta \Phi(t)=\Phi(t)-\langle\Phi\rangle$, with the angular brackets corresponding to thermal average $\langle
A\rangle=\Tr{\{\rho_T A\}}$, with $\rho_T=\textrm{e}^{-{\cal
    H}/(k_BT)}/Z$ the thermal equilibrium density matrix, ${\cal
  H}$ the spin Hamiltonian, and $Z=\Tr{\{\textrm{e}^{-{\cal H}/(k_BT)}\}}$ the partition function.  
       
We perform a spectral decomposition of 
Eq.~(\ref{stildedef}) by formally diagonalizing the spin Hamiltonian, 
${\cal H}|\alpha\rangle=E_{\alpha}|\alpha\rangle$:
\begin{equation}
\tilde{S}_{\Phi}(f)=\sum_{\alpha,\beta}\frac{\textrm{e}^{-\frac{E_{\alpha}}{k_BT}}}{Z}
\left|\langle\alpha|\delta\Phi(0)|\beta\rangle\right|^{2}\delta\left(f-\frac{E_{\beta}-E_{\alpha}}{h}\right).
\label{specdecomp}
\end{equation}
Any mechanism that couples the spins to the lattice\cite{desousa07} or
to themselves\cite{faoro08,lanting14,atalaya14} leads to finite frequency noise. 
However, evaluating Eq.~(\ref{specdecomp}) for a large spin system is a
challenging task that requires a series of uncontrolled approximations. 

The mechanism of frequency dependence in Eq.~(\ref{specdecomp}) is a
subject of current research.  In Ref.~\onlinecite{lanting14} the
frequency dependence of Eq.~(\ref{specdecomp}) was evaluated under the
assumption that the spins are in a paramagnetic phase with spin
dynamics governed by spin diffusion, leading to $1/f^{\alpha}$ noise
with exponent $0<\alpha < 1.5$ dependent on frequency range and device geometry.
Reference~\onlinecite{atalaya14} proposed a model based on spin-clusters
and non-diffusive dynamics driven by spin-spin dipolar interaction. It
was argued that the distribution of cluster sizes gives rise to a
large spread of spin-flip times and $1/f^{0.85}$ noise over a broad
frequency range, independent of device geometry.

Here we shall focus our discussion on the total flux noise power,
\begin{eqnarray}
\left\langle \left(\delta\Phi\right)^{2}\right\rangle&=& \int_{-\infty}^{\infty}\tilde{S}_{\Phi}(f)df\nonumber\\
&=&\sum_{i,j,\alpha,\beta}F_{i\alpha}F_{j\beta}\bigg(\frac{1}{2}\left\langle\left\{s_{i\alpha},s_{j\beta}\right\}\right\rangle\nonumber\\
&&-\langle s_{i\alpha}\rangle\langle s_{j\beta}\rangle\bigg),
\label{powergen}
\end{eqnarray}
where $\{A,B\}=AB+BA$ denotes the anticommutator of two operators. We note that Eq.~(\ref{powergen}) is independent on the particular
interaction mechanism driving spin dynamics; thus it allows a model
independent comparison of the role of $\bm{F}$ and the contribution arising from different spin
species. We take a continuum limit by introducing the magnetization or spin density
\begin{equation}
\bm{M}(\bm{r})=-\sum_{i}\bm{s}_i \delta(\bm{r}-\bm{R}_i),
\label{defmag}
\end{equation}
and defining the spin-spin spatial correlation function as
\begin{equation}
  C^{\alpha\beta}(\bm{r},\bm{r}')=\frac{1}{2}\left\langle \{M_{\alpha}(\bm{r}),M_{\beta}(\bm{r}')\}\right\rangle 
- \left\langle M_{\alpha}(\bm{r})\right\rangle\left\langle M_{\beta}(\bm{r}')\right\rangle.
\label{cab}
\end{equation}
In the continuum limit Eq.~(\ref{powergen}) becomes
\begin{equation}
\left\langle \left(\delta\Phi\right)^{2}\right\rangle = \int d^{d} r\int d^{d}r' \sum_{\alpha,\beta}F_{\alpha}(\bm{r})C^{\alpha\beta}(\bm{r},\bm{r}')F_{\beta}(\bm{r}'),
\label{powercont}
\end{equation}
with dimension $d=2$ (surface) and $d=3$ (bulk) to be considered below. 

The noise power Eq.~(\ref{powercont}) depends on spin texture through
the spatial correlation function $C^{\alpha\beta}(\bm{r},\bm{r}')$.
If the spin system is close to a phase transition, noise power
will show strong temperature dependence because of the formation of
spin clusters, which are described in general by deviations of
$C^{\alpha\beta}(\bm{r},\bm{r}')$ from a delta function $\delta(\bm{r}-\bm{r}')$. 
In addition, Eq.~(\ref{powercont}) may be temperature dependendent even in the absence of spin clusters, as we show below. 

\subsection{Noise power without spatial correlation (independent spins)\label{subsec:theorywithoutcorr}}

There are many situations where the correlation function is well
approximated by a delta function, and the state of any spin is independent of the others. These include high temperature, spin
textures with zero correlation (e.g. spin glass), and when all spins are fully polarized along one direction. Below we consider each case in detail.

In the limit of high temperature, defined by $k_BT$ being much larger than any energy scale affecting the spins, we may carry out a $1/T$ expansion of
Eq.~(\ref{cab}) and retain only the leading order contribution (zeroth
power of $1/T$, which is exact at $T=\infty$). This leading order contribution has
all different spin configurations occurring with equal probability,
implying $\langle s_{i\alpha}\rangle=0$ and $\langle
s_{i\alpha}s_{j\beta}\rangle =
\frac{S(S+1)}{3}\delta_{ij}\delta_{\alpha\beta}$. Further assuming that the spins are uniformly distributed in space with density $\sigma_d$ leads to
\begin{equation}
C^{\alpha\beta}_{T=\infty}(\bm{r},\bm{r}')=\frac{S(S+1)}{3}\sigma_d \delta(\bm{r}-\bm{r}')\delta_{\alpha\beta},
\label{chighT}
\end{equation}
and the high temperature noise power
\begin{equation}
\left\langle \left(\delta\Phi\right)^{2}\right\rangle_{T=\infty}= \frac{S(S+1)}{3}\sigma_d \int d^{d}r \left|\bm{F}(\bm{r})\right|^{2},
\label{noisepowerinfT}
\end{equation}
which is formally exact at $T=\infty$. In the absence of an external
magnetic field, Eq.~(\ref{noisepowerinfT}) is a good approximation for
nuclear spins down to $\mu$K temperatures (the energy scale for
nuclear-nuclear dipolar interaction between nearest-neighbor nuclear
spins).

If the spin system is in a phase that has approximately zero spatial correlation ($\langle s_{i\alpha}s_{j\beta}\rangle
\approx \langle s_{i\alpha}\rangle \langle s_{j\beta}\rangle$ for $i\neq j$), and if they are uniformly distributed in space,
the correlation function may be approximated by
\begin{subequations}
\begin{eqnarray}
C^{\alpha\beta}_{{\rm unc.}}(\bm{r},\bm{r}')&=& f^{\alpha\beta}(\bm{r},T)\sigma_d \delta(\bm{r}-\bm{r}'),\label{uncorr}\\
f^{\alpha\beta}(\bm{R}_i,T)&=&\left[\frac{1}{2}\langle \{s_{i\alpha},s_{i\beta}\}\rangle-\langle s_{i\alpha}\rangle \langle s_{i\beta}\rangle\right],\label{fab}
\end{eqnarray}
\end{subequations}
leading to the uncorrelated spin noise power
\begin{equation}
\left\langle \left(\delta\Phi\right)^{2}\right\rangle_{{\rm unc.}}= \sigma_d \int d^{d}r \sum_{\alpha,\beta}F_{\alpha}(\bm{r})f^{\alpha\beta}(\bm{r},T)F_{\beta}(\bm{r}).
\label{noisepowerunc}
\end{equation}
This expression is a good approximation in at least three cases of
interest: (1) At temperatures higher than the spin-spin coupling $J$,
but lower than single-spin anisotropy energy (which is non-zero for
$S>1/2$).  Assuming the single-spin anisotropy is equal for all spins,
the function $f^{\alpha\beta}(\bm{r},T)$ will depend only on
temperature and not on $\bm{r}$, and will differ from
$\delta_{\alpha\beta}$ signaling the presence of anisotropy. (2) At
all temperatures, when the spin-spin coupling alternates in sign
randomly, such as in a spin-glass.  As a result, the coarse-grained
spin-spin correlation function will average out to zero over the
length scales of $F_{\alpha}(\bm{r})$, but the long time averages
$\langle s_{i\alpha}\rangle$ will remain non-zero (non-ergodicity). A
space-dependent function $f^{\alpha\beta}(\bm{r},T)$ may be used to
model the lack of translation symmetry of the spin state. In the
simpler case of translation symmetry, a spin-diffusion model with
$f^{\alpha\beta}(T)=S(S+1)\chi(T)/\left(3\chi_0\right)\delta_{\alpha\beta}$
[Eq.~(9) of Ref.~\onlinecite{lanting14}] was proposed to describe the
frequency and temperature dependence of the spin-glass noise.  (3)
When all spins are polarized along the same direction, which occurs at
low $T$ in a ferromagnetic phase, or in the presence of a large
external magnetic field.

To illustrate case (1), consider a model spin Hamiltonian with easy-axis anisotropy along $\bm{\hat{e}}_{\parallel}$, 
${\cal H}_{{\rm ani.}}=-K\sum_i s_{i\parallel}^{2}$, with $s_{i\parallel}=\bm{s}_i\cdot\bm{\hat{e}}_{\parallel}$ and $K>0$ the anisotropy energy. When $J\ll k_BT\ll K$ the spins will be equally distributed in the $\pm S$ eigenstates of $s_{i\parallel}$; after a simple calculation we get $f^{\parallel,\parallel}=\langle s_{i\parallel}^{2}\rangle=S^2$ and $f^{\perp 1,\perp 1}=f^{\perp 2,\perp 2}=\langle s_{i\perp 1}^{2}\rangle = S/2$, with all other correlations equal to zero, leading to
\begin{equation}
\left\langle \left(\delta\Phi\right)^{2}\right\rangle_{J\ll k_BT\ll K}\approx  \sigma_d \int d^{d}r \left[S^2 F_{\parallel}^{2}(\bm{r})+\frac{S}{2}F_{\perp}^{2}(\bm{r})\right].
\label{noiseanis}
\end{equation}
Depending on the direction of $\bm{F}(\bm{r})$, the noise power may
get reduced or increased in comparison to the high $T$ result
Eq.~(\ref{noisepowerinfT}).

More generally, the Hermitian matrix $f^{\alpha\beta}(\bm{r},T)$ can be
diagonalized to find its eigenvectors $\bm{\hat{f}}_\gamma$ and
eigenvalues
$a_{\gamma}=\langle(\bm{s}_i\cdot\bm{\hat{f}}_{\gamma})^{2}\rangle-\langle(\bm{s}_i\cdot\bm{\hat{f}}_{\gamma})\rangle^{2}\ge
0$.  We may establish a general inequality for uncorrelated noise
by noting that the largest eigenvalue of $f^{\alpha\beta}$ is smaller
than $S(S+1)$ (the eigenvalue of $\bm{s}_{i}^{2}$):
\begin{eqnarray}
\left\langle \left(\delta\Phi\right)^{2}\right\rangle_{{\rm unc.}}&=& \sigma_d \int d^{d}r \sum_{\gamma}a_{\gamma}\left|\bm{F}(\bm{r})\cdot \bm{\hat{f}}_{\gamma}\right|^{2}\nonumber\\
&\le&3\left\langle \left(\delta\Phi\right)^{2}\right\rangle_{T=\infty}.
\label{noisepoweruncorrineq}
\end{eqnarray}
If $S=1/2$ and the spin state has translation symmetry, $a_{\gamma}$
will be equal to $1/4$ for all $\gamma$, and the noise power is simply
equal to the $T=\infty$ result Eq.~(\ref{noisepowerinfT}). 

Finally, we consider case (3) in detail. At low $T$, the spins may spontaneously polarize if the spin-spin interaction is ferromagnetic (See section~\ref{subsec:withcorr}); alternatively 
they can be polarized with a strong external magnetic field. Measuring flux noise in the presence of a strong $B$
field is a quite challenging experiment. For superconductivity to
remain unnaffected, the external $B$ field has to be applied along the
plane of all wire segments, with the magnetic length remaining larger
than the thin-film width $b$. Moreover, any modulation of the critical current due to the presence of flux perpendicular to the Josephson junctions will have to be accounted for. Despite
these challenges, we shall show that it is worth considering this
experiment because it would provide a mapping of $\bm{F}(\bm{r})$ and
the measurement of spin quantum number $S$.

When the Zeeman energy scale dominates, the spin Hamiltonian can be
approximated by ${\cal H}_{{\rm Zeeman}}=-\sum_i \bm{\mu}_i\cdot
\bm{B}$, with $\bm{\mu}_i=-g\mu_s \bm{s}_i$.  Here we use
$\mu_s=\mu_B$ for electron spins ($\mu_B$ is the Bohr magneton), and
$\mu_s=-\mu_N$ for nuclear spins ($\mu_N$ is the nuclear magneton),
with $g$ the $g$-factor ($g$ is quite close to $2$ for most electron
impurities but can be very different than $2$ for nuclei). In this
regime correlation between different spins will be zero and the
functions $f^{\alpha\beta}(\bm{r},T)$ in Eq.~(\ref{fab}) can be
separated into two kinds, $f_{\parallel}(T)$ and $f_{\perp}(T)$:
\begin{subequations}
\begin{eqnarray}
f_{\parallel}(\tilde{B})&=&\langle (\bm{s}_i\cdot \bm{\hat{B}})^{2}\rangle -\langle(\bm{s}_i\cdot \bm{\hat{B}}) \rangle^{2}\nonumber\\
&=&\frac{1}{4}\left\{ \frac{1}{\sinh^{2}{(\tilde{B})}}-\frac{(2S+1)^{2}}{\sinh^{2}{[(2S+1)\tilde{B}]}}\right\},\\
f_{\perp}(\tilde{B})&=& \langle (\bm{s}_i\cdot\bm{\hat{B}}_{\perp 1})^{2} \rangle=\langle(\bm{s}_i\cdot \bm{\hat{B}}_{\perp 2})^{2} \rangle\nonumber\\
&=& \frac{1}{4} \coth{(\tilde{B})} \left\{(2S+1) \coth{[(2S+1)\tilde{B}]}\right.\nonumber\\
&&\left.-\coth{(\tilde{B})}\right\},
\end{eqnarray}
\end{subequations}
where $\bm{\hat{B}}$ is the unit vector along the magnetic field, $\bm{\hat{B}}_{\perp 1}, \bm{\hat{B}}_{\perp 2}$ is a set of orthogonal unit vectors perpendicular to it, and $\tilde{B}=g\mu_s B/(2k_B T)$ measures the strength of the field. The equilibrium spin polarization is given by
\begin{equation}
\langle \bm{s}_i\rangle=-\frac{1}{2}\{(2S+1)\coth{[(2S+1)\tilde{B}]}-\coth{(\tilde{B})}\}\bm{\hat{B}}.
\end{equation}
Plugging these expressions into Eq.~(\ref{noisepowerunc}) yields
\begin{eqnarray}
\left\langle \left(\delta\Phi\right)^{2}\right\rangle_{{\rm High}\;B} &\approx& \sigma_d \int d^d r \left[f_{\perp}\left|\bm{F}(\bm{r})\right|^{2}\right.\nonumber\\
&&\left.-\left(f_{\perp}-f_{\parallel}\right)\left|\bm{F}(\bm{r}) \cdot \bm{\hat{B}}\right|^{2}\right].
\label{powerB}
\end{eqnarray}
Since $f_{\perp}(\tilde{B})>f_{\parallel}(\tilde{B})$ for all $\tilde{B}$, Eq.~(\ref{powerB}) shows that the noise power \emph{gets reduced upon the application of external $B$ field}. In the limit $\tilde{B}\gg 1$, all spins are polarized leading to
\begin{equation}
\left\langle \left(\delta\Phi\right)^{2}\right\rangle_{{\rm pol.}\parallel\bm{\hat{B}}} = \frac{S}{2} \sigma_d\int d^{d}r \left[\left|\bm{F}(\bm{r})\right|^{2}-\left|\bm{F}(\bm{r}) \cdot \bm{\hat{B}}\right|^{2}\right],
\label{powerBlimit}
\end{equation}
which is quite different from the high $T$ result
Eq.~(\ref{noisepowerinfT}). Equation~(\ref{powerBlimit}) is the \emph{exact noise power of a fully polarized spin system}, irrespective of whether the polarization occurs because of application of a $B$ field, or due to spontaneous symmetry breaking e.g. in a ferromagnetic ground state. 
Thus, measurements of the dependence of
noise as a function of the direction and magnitude of the in-plane
magnetic field, or on direction of the spin polarization provides information on the components of the flux
vector $\bm{F}(\bm{r})$ and on the value of the spin quantum number
$S$.
 
As we show below, the direction of $\bm{F}(\bm{r})$ is poloidal along the
surface of the wire (Fig.~\ref{FigF}), implying that \emph{spin polarization along a single direction} can never completely suppress flux noise.  Note
how the residual noise arises due to the quantum nature of the spins. For example,
if the spins are polarized along $\bm{\hat{x}}$, they have equal probability of pointing
along any of the four directions
$+\bm{\hat{y}},-\bm{\hat{y}},+\bm{\hat{z}},-\bm{\hat{z}}$. This
uncertainty gives rise to quantum noise power proportional to
$F_{iy}^{2}+F_{iz}^{2}$.  

Finally, consider a spin-polarized state where each spin located at $\bm{R}_i$ points parallel or antiparallel to $\bm{F}_i$; we call this the poloidal state.The noise power produced by this state is obtained by 
evaluating the average in Eq.~(\ref{powergen}) with a single product state of $N$ spins each pointing along $\bm{F}_i$; the result is identical to Eq.~(\ref{powerBlimit}) with 
$\bm{\hat{B}}\rightarrow \bm{\hat{F}}(\bm{r})$, leading to
\begin{equation}
\left\langle \left(\delta\Phi\right)^{2}\right\rangle_{{\rm Poloidal}} =0.
\label{fnpoloidal}
\end{equation}
Therefore, engineering the poloidal spin texture enables complete suppression of flux noise. We emphasize that this is an exact result; it happens because in the poloidal state every single spin quantum fluctuation has to be perpendicular to $\bm{F}_i$, producing zero flux noise.

\begin{figure}
\includegraphics[width=3.4in]{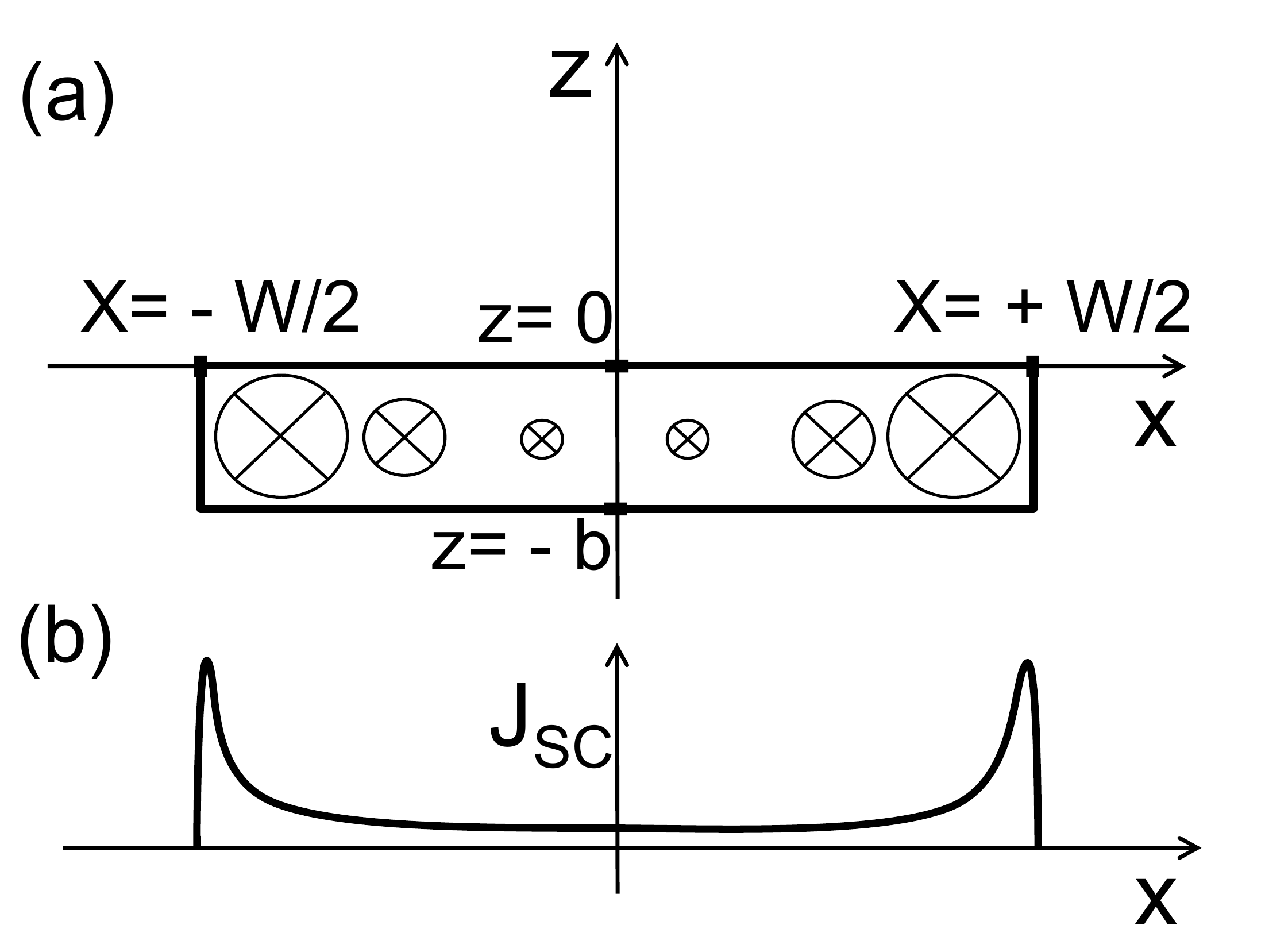}
\caption{(a) Transverse cross-section of the SQUID wire (See also Fig.~\ref{FigF}). 
The current density is shown pointing along the $y$-direction. (b) Magnitude of the superconductor current density as a function of $x$ [Eq.~(\ref{jsquid})].}
\label{Figcoord}
\end{figure}

\begin{figure*}
\centering
\subfigure[Spin pointing along $x$ (along the wire surface)]{\includegraphics[width=0.49\textwidth]{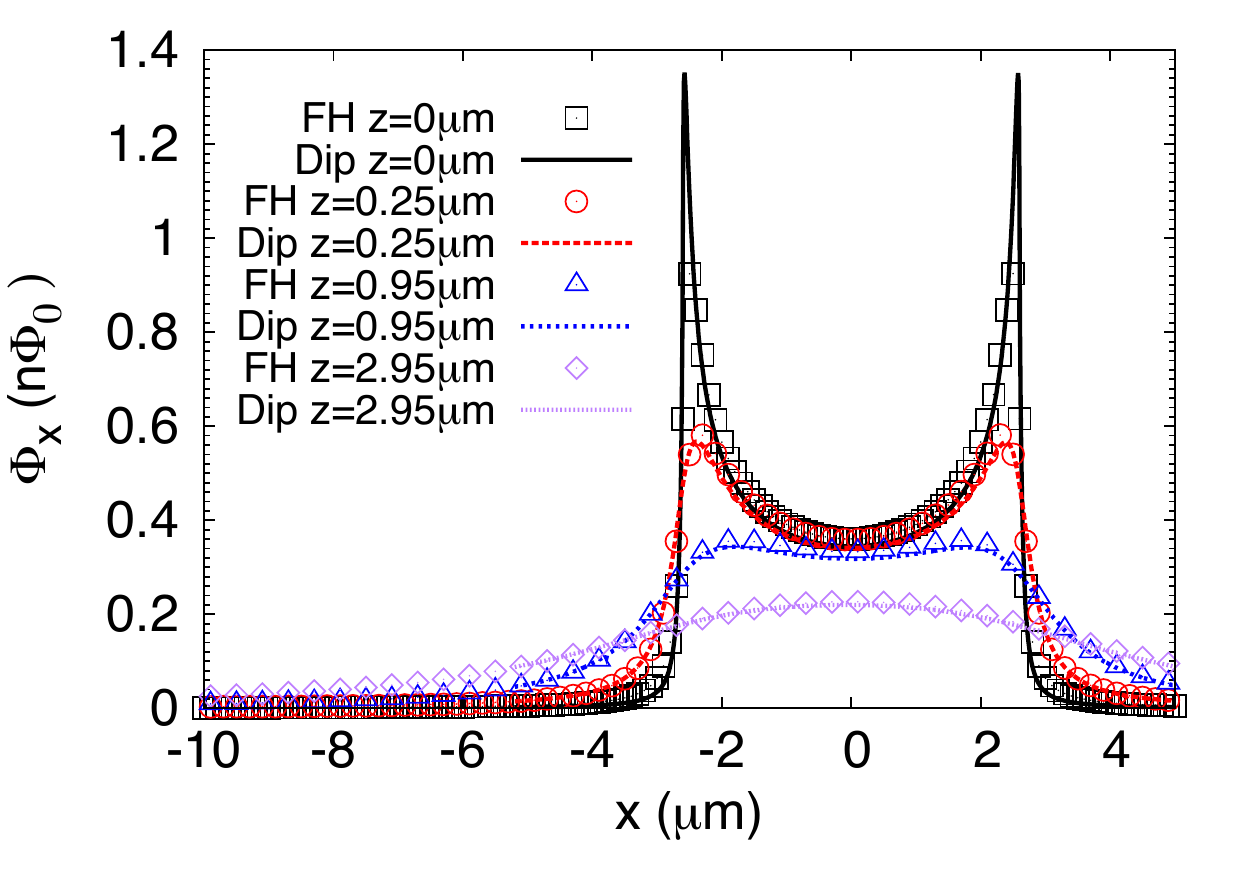} }
\subfigure[Spin pointing along $z$ (perpendicular to the wire surface)]{\includegraphics[width=0.49\textwidth]{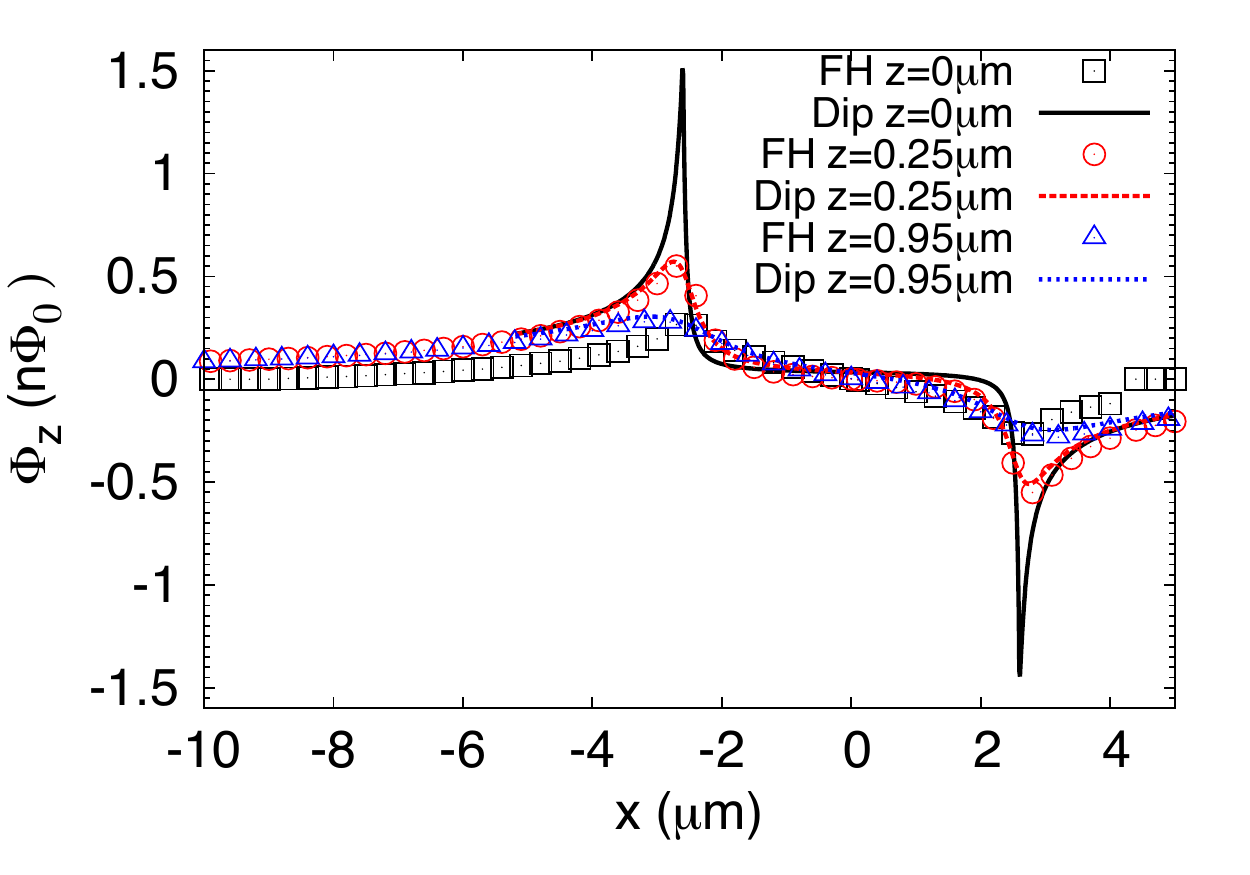}  }
\caption[Comparison of the Flux calculated by FastHenry and the numerical dipole method]{(color online). Comparison of the Flux produced by each spin as a function of spin location calculated by FastHenry and the numerical dipole method, for a superconducting wire of penetration depth $\lambda=0.07\;\mu$m, thickness $b=0.1\;\mu$m, and width $W=5.2\;\mu$m.  The coordinate $x$ runs along the lateral width of the wire, with edges at $x = \pm 2.6 \mu$m.}
\label{fig:compare}
\end{figure*}

\subsection{Noise power with spatial correlation (ferromagnetic spin clusters)\label{subsec:withcorr}}

The interaction between the spins may lead to non-zero spatial correlation. 
To see the impact on flux noise power, consider the Heisenberg model with easy-axis anisotropy,
\begin{equation}
{\cal H}_{{\rm s-s}}=-\frac{1}{2}\sum_{i,j}J(\bm{R}_i-\bm{R}_{j})\bm{s}_i\cdot\bm{s}_j -K\sum_i s_{i\parallel}^{2},
\label{heisenberg}
\end{equation}
where $K\geq 0$ and $s_{i\parallel}=\bm{s}_i\cdot\bm{\hat{e}}_{\parallel}$, with $\bm{\hat{e}}_{\parallel}$ the direction of the anisotropy axis. 
We take the continuum limit with
$\bm{M}(\bm{r})$ as in Eq.~(\ref{defmag}), and expand Eq.~(\ref{heisenberg}) up to the second order in spatial derivatives:
\begin{eqnarray}
{\cal H}_{{\rm s-s}}&\approx& \int d^{d} r \Bigg\{\frac{\bar{J}\xi_{0}^{d}}{2}\left[ -\left|\bm{M}\right|^{2}+\xi_{0}^{2}\sum_{\alpha=x,y,z}\left|\nabla M_{\alpha}\right|^{2}\right]\nonumber\\
&&-\frac{K}{\sigma_d}M_{\parallel}^{2}\Bigg\}.
\label{hexpanded}
\end{eqnarray}
This expression is valid in the long-wavelength approximation, when terms of order 
$|\nabla^{2}M|^{2}$ are negligible.

The exchange interaction is assumed to satisfy
\begin{equation}
\int d^{d}r J(\bm{r})x_{\alpha}x_{\beta} = 2\xi_{0}^{d+2}\bar{J}\delta_{\alpha\beta},
\end{equation}
with parameters $\xi_0$ and  $\bar{J}$  defined by
\begin{subequations}
\begin{eqnarray}
\xi_{0}^{2}&=&\frac{1}{2}\frac{\int d^{d}r J(\bm{r})x_{\alpha}^{2}}{\int d^{d}r J(\bm{r})},\\
\bar{J}&=&\frac{1}{\xi_{0}^{d}}\int d^{d}r J(\bm{r}).
\end{eqnarray}
\end{subequations}
Parameter $\xi_0$ models the range of the exchange interaction, while
$\bar{J}$ models its average strength.  Many competing interactions
contribute to $J(\bm{r})$, such as direct exchange (always
antiferromagnetic) and RKKY (ferromagnetic when
$k_F|\bm{R}_i-\bm{R}_j|\ll 1$).\cite{faoro08,atalaya14,de14} We shall
focus in the case $0<\bar{J}<\infty$ and $\xi_0<\infty$, when the
system is able to transition into a ferromagnetic state with $\bm{M}\parallel \pm\bm{\hat{e}}_{\parallel}$. We shall also use  Eq.~(\ref{hexpanded}) with a space-dependent $\bm{\hat{e}}_{\parallel}=\bm{\hat{F}}(\bm{r})$ as a toy model that we call the \emph{poloidal model}. When $K\gg \bar{J}\sigma_d \xi_{0}^{d} (\xi_0/b)^{2}$ such a model has the poloidal state $\bm{M}(\bm{r})\parallel\pm \bm{F}(\bm{r})$ as its $T=0$ ground state.

We use mean-field theory;\cite{chaikin00} this is
done by subtracting $T$ times the single spin entropy from
Eq.~(\ref{hexpanded}). The result is that the system develops
a ferromagnetic moment $\langle M_{\parallel}\rangle\neq 0$ at the critical temperature
\begin{equation}
k_BT_c=\frac{S(S+1)}{3}\left(\sigma_d \xi_{0}^{d}\bar{J}+2K\right),
\label{tcfm}
\end{equation}
with correlation functions given by:
\begin{subequations}
\begin{eqnarray}
C^{\parallel,\parallel}(\bm{r},\bm{r}')&=&\frac{k_BT}{\bar{J}\xi_{0}^{d+2}}
\int \frac{d^{d}q}{(2\pi)^{d}}\frac{\textrm{e}^{i\bm{q}\cdot (\bm{r}-\bm{r}')}\xi^{2}_{\parallel}}{1+\left(\xi_{\parallel} q\right)^{2}},\label{mfcpar}
\\
C^{\perp \alpha,\perp\beta}(\bm{r},\bm{r}')&=&\frac{k_BT}{\bar{J}\xi_{0}^{d+2}}
\int \frac{d^{d}q}{(2\pi)^{d}} \frac{\textrm{e}^{i\bm{q}\cdot (\bm{r}-\bm{r}')}\xi_{\perp}^{2}}{1+\left(\xi_{\perp} q\right)^{2}}\nonumber\\
&&\times\delta_{\alpha\beta},\label{mfcperp}
\end{eqnarray}
\end{subequations}
where $\perp\alpha$ with $\alpha=1,2$ denote the two directions perpendicular to $\bm{\hat{e}}_{\parallel}$ (the correlation function between $\parallel$ and $\perp\alpha$ is zero). 
There are two correlation length scales $\xi_{\parallel}$ and $\xi_{\perp}$:
\begin{subequations}
\begin{eqnarray}
\xi_{\parallel}(T)&=&\xi_0 \sqrt{\frac{T'}{u_T|T-T_c|}},\label{xifma}\\
\xi_{\perp}(T>T_c)&=&\xi_0 \sqrt{\frac{T'}{\left(T-T'\right)}},\label{xiperpa}\\
\xi_{\perp}(T<T_c)&=&\xi_0 \sqrt{\frac{\sigma_d \xi_{0}^{d}\bar{J}}{2K}},\label{xiperpb}
\end{eqnarray}
\end{subequations}
where $k_BT'=[S(S+1)/3]\sigma_d \xi_{0}^{d}\bar{J}$ and $u_T=1$ for $T>
T_c$ and $u_T=2$ for $T<T_c$.  The length scales
$\xi_{\parallel},\xi_{\perp}$ describe the average size for spin
clusters polarized along $\bm{\hat{e}}_{\parallel}$, and the direction
perpendicular to it, respectively. Note $T_c>T'$, so that only
$\xi_{\parallel}$ diverges at the transition.
For $d=2$ we get 
\begin{subequations}
\begin{eqnarray}
C^{\parallel,\parallel}_{d=2}(\bm{r}-\bm{r}')&=&\frac{k_BT}{2\pi\bar{J}\xi_{0}^{4}}K_{0}\left(\frac{|\bm{r}-\bm{r}'|}{\xi_{\parallel}}\right),\label{cpar2d}\\
C^{\perp\alpha,\perp\beta}_{d=2}(\bm{r}-\bm{r}')&=&\frac{k_BT}{2\pi\bar{J}\xi_{0}^{4}}K_{0}\left(\frac{|\bm{r}-\bm{r}'|}{\xi_{\perp}}\right)\delta_{\alpha\beta},\label{cperp2d}
\end{eqnarray}
\end{subequations}
where $K_{0}(x)$ is the modified Bessel function of the 2nd kind. 

The noise power for surface spins ($d=2$) in the presence of correlations is obtained by plugging Eqs.~(\ref{cpar2d}),~(\ref{cperp2d}) into Eq.~(\ref{powercont}), leading to
\begin{widetext}
\begin{equation}
\frac{\left\langle \left(\delta\Phi\right)^{2}\right\rangle_{{\rm corr.}}}{\left\langle \left(\delta\Phi\right)^{2}\right\rangle_{T=\infty}}
=\frac{T}{u_T|T-T_c|}
\frac{\int d^{2}r \int d^{2}r' \left[F_{\parallel}(\bm{r})F_{\parallel}(\bm{r}')K_0\left(\frac{|\bm{r}-\bm{r}'|}{\xi_{\parallel}}\right)+\bm{F}_{\perp}(\bm{r})\cdot \bm{F}_{\perp}(\bm{r}')K_0\left(\frac{|\bm{r}-\bm{r}'|}{\xi_{\perp}}\right)\right]}
{2\pi \xi_{\parallel}^{2}\int d^{2}r |\bm{F}(\bm{r})|^{2}},
\label{corrnoise2d}
\end{equation}
\end{widetext}
where  we presented correlated noise divided by the $T=\infty$
uncorrelated noise [Eq.~(\ref{noisepowerinfT})] for convenience.
Notably, Eq.~(\ref{corrnoise2d}) depends on model parameters only
through $\xi_{\parallel},\xi_{\perp}$ and $T_c$.  Once again, the vector nature of
$\bm{F}(\bm{r})$ plays an important role in determining noise power: Correlated noise
tends to decrease (increase) when $\bm{F}(\bm{r})$ and
$\bm{F}(\bm{r}')$ are antiparallel (parallel) for $\bm{r},\bm{r}'$ in
different surfaces of the wire (See e.g. the top and bottom wire
surfaces in Fig.~\ref{FigF}).

We remark that mean-field theory neglects critical fluctuations and is
not necessarily a good approximation for $T$ close to $T_c$. For
$d=1,2$ and $K=0$ (or $S=1/2$, when easy-axis anisotropy is effectively zero),
critical fluctuations reduce the actual $T_c$ of the
Heisenberg model to zero.\cite{mermin66} 
For $K>0$, $T_c$ is generally non-zero at $d=2$, but can be
substantially reduced in comparison to the mean-field prediction
Eq.~(\ref{tcfm}). However, \emph{mean-field
  theory is an excellent approximation in the region $T\gg T_c$, when
  critical fluctuations play no role}. To see this, consider the $T\gg
T_c$ limit of Eqs.~(\ref{xifma})--(\ref{cperp2d}): We get $\xi_{\parallel},\xi_{\perp}\approx
\xi_0\sqrt{T'/T}\rightarrow 0$, and $K_0(|\bm{r}-\bm{r}'|/\xi)\rightarrow
2\pi \xi^{2}\delta(\bm{r}-\bm{r}')$, leading to
$C^{\alpha\beta}(\bm{r}-\bm{r}')\approx
k_BT'/(\bar{J}\xi_{0}^{2})\delta(\bm{r}-\bm{r}')\delta_{\alpha\beta}$.
Plugging $k_BT'=[S(S+1)/3]\sigma_d \xi_{0}^{d}\bar{J}$ we see that this is identical to the
$T=\infty$ limit shown in Eq.~(\ref{chighT}). 

Furthermore, mean-field theory neglects quantum fluctuations; this is
a problem for the $T\ll T_c$ region, where mean-field theory is exact
only in the limit $S\rightarrow \infty$. When $J(\bm{R}_i-\bm{R}_j)>0$
for all $i,j$, the exact $T=0$ state of the
system will have all spins fully polarized along either
$+\bm{\hat{e}}_{\parallel}$ or $-\bm{\hat{e}}_{\parallel}$ (or an arbitrary direction for $S=1/2$) leading to
\begin{subequations}
\begin{eqnarray}
C^{\parallel,\parallel}_{T=0}(\bm{r},\bm{r}')&=&0,\label{cparpol}\\
C^{\perp\alpha,\perp\beta}_{T=0}(\bm{r},\bm{r}')&=&\frac{S}{2}\sigma_d \delta(\bm{r}-\bm{r}')\delta_{\alpha\beta}.\label{cperppol}
\end{eqnarray}
\end{subequations}
Hence $C^{\perp\alpha,\perp\alpha}(\bm{r},\bm{r}')$ remains non-zero when $T\rightarrow 0$,
signaling the presence of quantum fluctuations. Plugging
Eqs.~(\ref{cparpol}),~(\ref{cperppol}) into Eq.~(\ref{powercont}) we
get the spin polarized noise Eq.~(\ref{powerBlimit}) with
$\hat{\bm{B}}=\bm{\hat{e}}_{\parallel}$. While this result disagrees with
mean-field theory, we note that the ratio of Eq.~(\ref{powerBlimit}) to
Eq.~(\ref{noisepowerinfT}) is proportional to $1/(S+1)$. When
$S\rightarrow \infty$ this ratio goes to zero, in agreement with the
$T\rightarrow 0$ limit of Eq.~(\ref{corrnoise2d}).

\section{Evaluation of the flux vector: The numerical dipole method \label{sec:numermeth}}

When an electron travels across a closed path ${\cal C}$ inside the SC wire, it is affected by a flux equal to $\Phi_i = \int_{{\cal C}}\bm{A}_i(\bm{r})\cdot d\bm{l}$, where $\bm{A}_i(\bm{r})$ is the vector potential for the spin $\bm{s}_i$ located at $\bm{R}_i$. For an extended wire with SC current density $\bm{J}_{{\rm SC}}(\bm{r})$ and total current $I_{{\rm SC}}$, the corresponding flux can be obtained by representing the wire by a set of infinitesimally thin closed paths ${\cal C}$, and summing over all the paths with $d\bm{l}=d^{3}r\bm{J}_{{\rm SC}}(\bm{r})/I_{{\rm SC}}$, a weighting factor that represents 
the fraction of current flowing through each path:
\begin{equation}
\Phi_i=\int d^{3}r \bm{A}_{i}(\bm{r})\cdot \bm{J}_{{\rm SC}}(\bm{r}),
\label{phiigen}
\end{equation}
where the integral is over the region where $\bm{J}_{{\rm SC}}\neq 0$, i.e. the 
volume of the wire. 
The numerical dipole method
consists in using the spin-dipole expression for $\bm{A}_i(\bm{r})$,\cite{jackson99}
\begin{equation}
\bm{A}_i(\bm{r})= -\frac{g\mu_s\mu_0}{4\pi} \frac{\bm{s}_i \times\left(\bm{r}-\bm{R}_i\right)}{\left|\bm{r}-\bm{R}_i\right|^{3}},
\label{ai}
\end{equation}
together with an analytical approximation for $\bm{J}_{SC}(\bm{r})$. 
Plugging Eq.~(\ref{ai}) into Eq.~(\ref{phiigen}) we get $\Phi_i=-\bm{F}(\bm{R}_i)\cdot \bm{s}_i$ with
\begin{subequations}
\begin{eqnarray}
\bm{F}(\bm{R})&=&\frac{g\mu_s\mu_0}{4\pi} \int d^3 r \frac{(\bm{r}-\bm{R})
\times \bm{J}_{\rm{SC}}(\bm{r})}{\left|\bm{r}-\bm{R}\right|^{3}I_{\rm{SC}}}\label{fvec1}\\
&=& \frac{g\mu_s\mu_0}{4\pi}\nabla_{\bm{R}}\times\left[\int d^3r \frac{\bm{J}_{{\rm SC}}(\bm{r})}{I_{{\rm SC}}\left|\bm{r}-\bm{R}\right|}\right].
\label{fvec2}
\end{eqnarray}
\end{subequations}
Note that we choose a sign convention so that $\bm{F}(\bm{R})$ points along $\bm{B}(\bm{R})$, the magnetic field produced by $\bm{J}_{{\rm SC}}$ at $\bm{R}$, and that
the direction of $\bm{F}$ can be found using the right hand rule with the thumb pointing towards $\bm{J}_{{\rm SC}}$ (Fig.~\ref{FigF}). 
The corresponding Zeeman energy of the spin system is given by ${\cal H}_{{\rm Zeeman}}=-I_{{\rm SC}}\int d^{d}r \bm{F}(\bm{r})\cdot \bm{M}(\bm{r})$. 
We remark that this same result can be obtained from the flux-inductance theorem.\cite{fluxinductancenote}

The current density $\bm{J}_{{\rm SC}}$ includes contributions from external sources plus screening currents causing Meissner effect. To find $\bm{J}_{{\rm SC}}$
one usually has to integrate the London equations numerically with the help of software packages such as FastHenry.\cite{smithhisler96}
However, there exists an important case where $\bm{J}_{{\rm SC}}$ is known analytically: 
Thin-film wires of width $b\lesssim \lambda$, where $\lambda$ is the SC penetration depth ($\lambda=0.05-0.1\;\mu$m for most superconductors), 
and wire lateral width $W$ that is large enough to satisfy 
\begin{equation}
\tilde{\lambda}\equiv \frac{\lambda^2}{bW}\ll 1,
\label{lambdatilde}
\end{equation}
and $\xi_{{\rm SC}}\ll W$, where $\xi_{{\rm SC}}$ is the
superconductor coherence length including the electron mean free
path (this latter condition ensures that the London equations with a
single length scale $\lambda$ is a good approximation).

Below we present analytic results in terms of powers of the small parameter $\tilde{\lambda}$. 
Neglecting terms that are first order in $\tilde{\lambda}$, the SC wire current density 
can be written as\cite{rhoderick62,vanduzer99}
\begin{widetext}
\begin{equation}
\bm{J}_{{\rm SC}}(\bm{r})=\frac{2I_{SC}}{\pi b W\left(1-\gamma \sqrt{2\tilde{\lambda}}\right)}\bm{\hat{y}}\left\{
\begin{array}{l l}
\frac{1}{\sqrt{2\tilde{\lambda}}}\textrm{e}^{-\frac{(1-\tilde{\lambda}-|u|)}{2\tilde{\lambda}}} & {\rm for}\;(1-\tilde{\lambda})<|u|\leq 1,\\
\frac{1}{\sqrt{1-u^{2}}} & {\rm for}\;|u|\leq (1-\tilde{\lambda}),
\end{array}
\right.
\label{jsquid}
\end{equation}
where $\gamma=2(2-\textrm{e}^{1/2})/\pi=0.2236$ is a numerical
constant, and the coordinate $u=2x/W$ runs along the lateral width of
the wire, as shown in Fig.~\ref{Figcoord}. The numerical dipole method
consists in using Eq.~(\ref{jsquid}) in Eq.~(\ref{fvec2}) to get an
approximation for the flux vector \emph{that neglects the feedback
  effect of the spins on $\bm{J}_{{\rm SC}}$}. This feedback effect
shall not be significant when the spins are unpolarized. Later we will
confirm this expectation by comparing our numerical dipole method to
exact integration of the London equations using FastHenry, and show that non-zero
spin polarization leads to an asymmetry on top of this solution.

\begin{figure*}
\centering
\subfigure[Spin pointing along $x$ (along the wire surface)]{\includegraphics[width=0.49\textwidth]{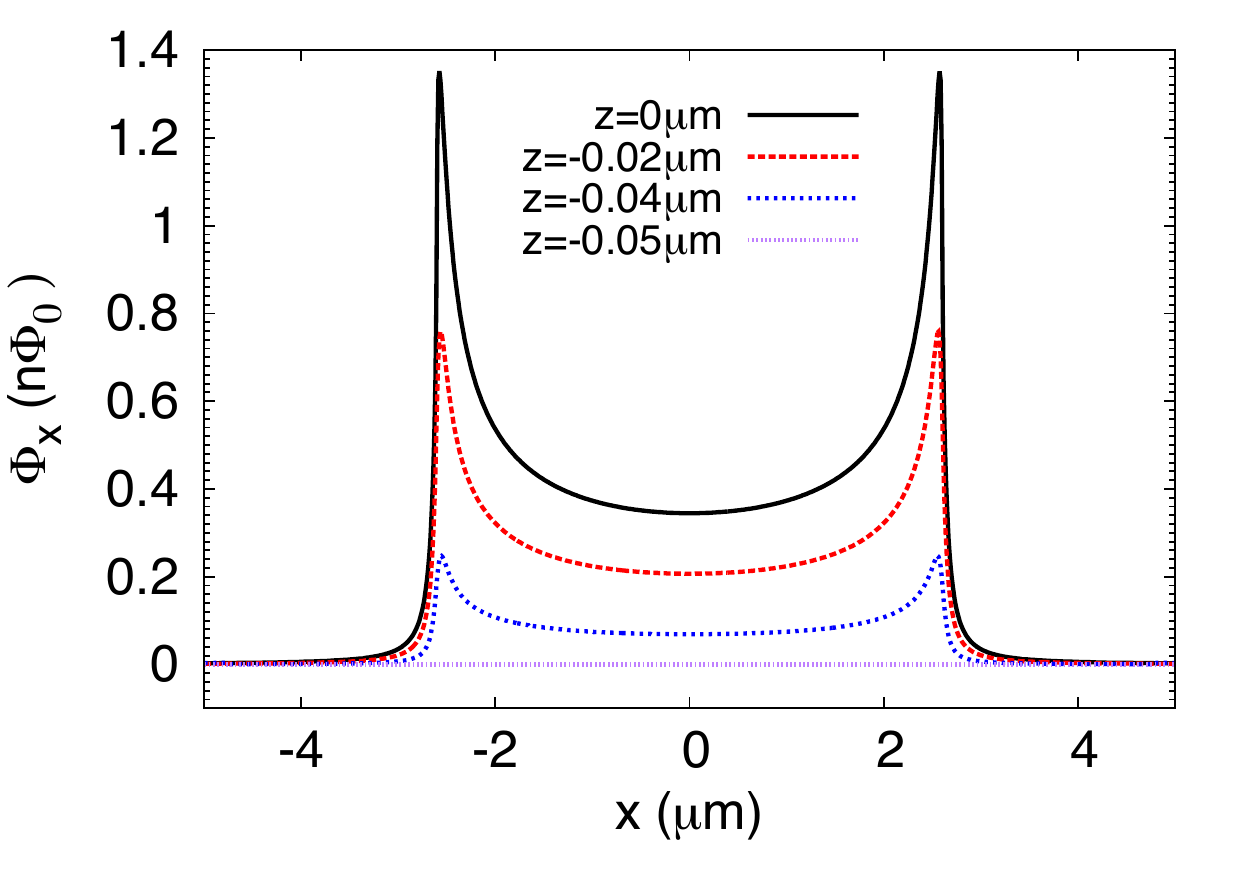} }
\subfigure[Spin pointing along $z$ (perpendicular to the wire surface)]{\includegraphics[width=0.49\textwidth]{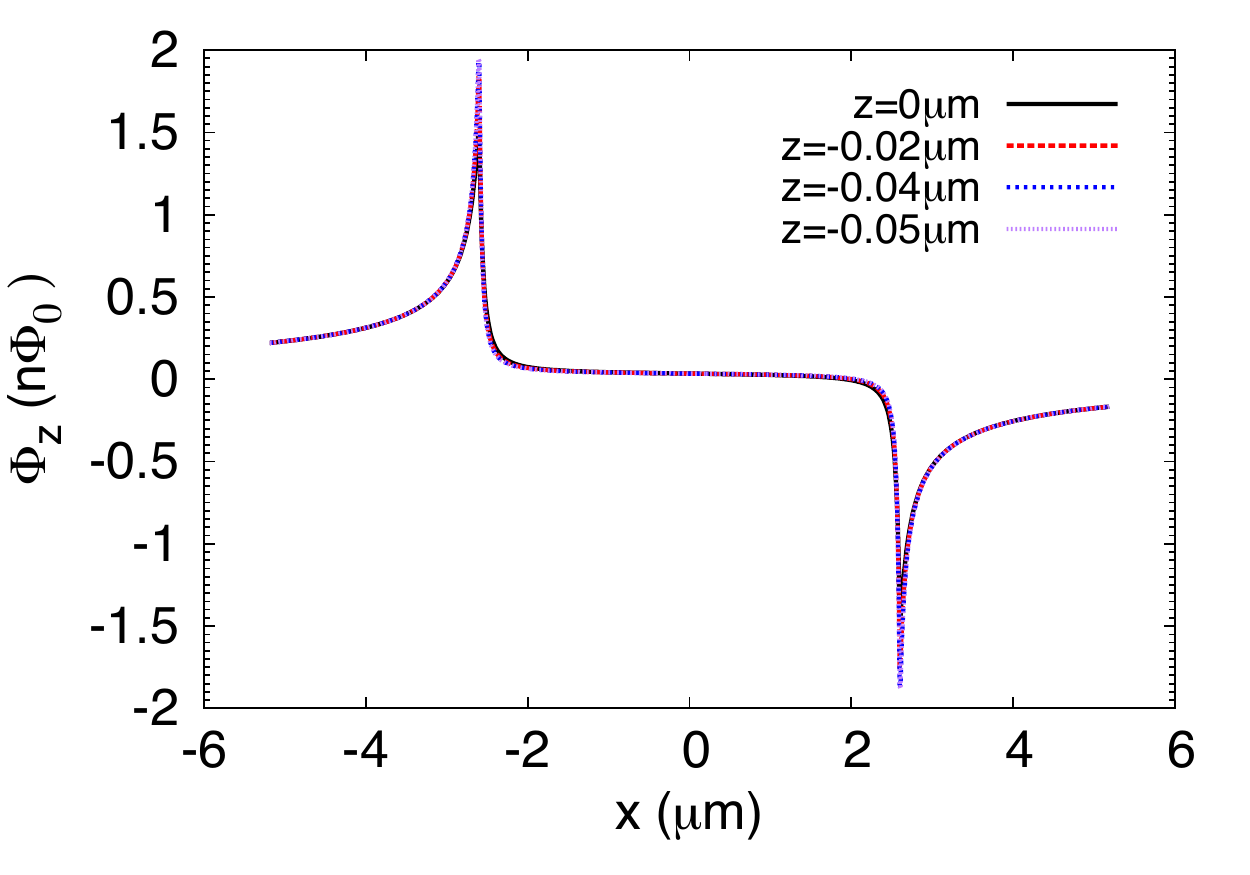}  }
\caption[Flux produced by each electron spin impurity as a function of
spin location calculated using the numerical dipole method for a
superconducting wire]{(color online). Flux produced by an electron spin impurity as a
  function of spin location inside the SC wire. All results used the
  numerical dipole method, for a superconducting wire of penetration depth $\lambda=0.07\;\mu$m, thickness $b=0.1\;\mu$m, and width $W=5.2\;\mu$m. The coordinate $x$ runs along the lateral width of
  the wire, with edges at $x = \pm 2.6 \mu$m. Note how the flux is zero in the region inside the wire
($z=-b/2$ and $x$ away from the edges).}
\label{fig:diponly}
\end{figure*}

It is straightforward to integrate Eq.~(\ref{fvec2}) explicitly for
the infinitely long wire with coordinate axes shown in Fig.~\ref{Figcoord}. For a
spin located at $\bm{R}=(X,0,Z)$,
\begin{subequations}
\begin{eqnarray}
F_x(X,Z)&=& -\frac{g\mu_s\mu_0W}{8\pi}\int_{-1}^{1}du \frac{J_{{\rm SC}}(u)}{I_{{\rm SC}}} 
\ln{\left[\frac{(x-X)^{2}+Z^{2}}{(x-X)^{2}+(Z+b)^{2}}\right]},\label{fxexp}\\
F_y(X,Z)&=&0,\label{fyexp}\\
F_z(X,Z)&=& \frac{g\mu_s\mu_0W}{4\pi}\int_{-1}^{1}du \frac{J_{{\rm SC}}(u)}{I_{{\rm SC}}}\left[
\arctan{\left(\frac{Z+b}{x-X}\right)}-
\arctan{\left(\frac{Z}{x-X}\right)}\right].\label{fzexp}
\end{eqnarray}
\end{subequations}

We can integrate a few particular cases 
analytically; neglecting terms that are first order in $\tilde{\lambda}$:
\begin{itemize}
\item \emph{Mid-surface, $(X=0, Z=0)$:}
\begin{subequations}
\begin{eqnarray}
F_x&\approx&\frac{g\mu_s\mu_0}{\pi W}\left[\frac{1-\gamma \frac{b}{W}\sqrt{2\tilde{\lambda}}}{1-\gamma\sqrt{2\tilde{\lambda}}}\right],\\
F_z&=&0.
\end{eqnarray}
\end{subequations}
This expression is a good approximation in the top surface away from the wire edge.
For electrons with $g=2$, and assuming $\sqrt{2\tilde{\lambda}}\ll 1$ we get
$|\bm{F}_{{\rm mid-surface}}|\approx \left(\frac{3.6\mu\textrm{m}}{W}\right)n\Phi_0$ ($n\Phi_0=10^{-9}\Phi_0$). 

\item \emph{Mid-edge, $(X=\frac{W}{2}, Z=-\frac{b}{2})$:} 
\begin{subequations}
\begin{eqnarray}
F_x&=&0,\\
F_z&\approx& -\frac{g\mu_s\mu_0}{\pi\sqrt{bW}\left(1-\gamma \sqrt{2\tilde{\lambda}}\right)}
\left[1-\gamma \frac{\lambda/b}{\sqrt{1+2\left(\frac{\lambda}{5b}\right)^{2}}}\right].\label{fmax}
\end{eqnarray}
\end{subequations}

\item \emph{Corner, $(X=\frac{W}{2}, Z=0)$:}
\begin{subequations}
\begin{eqnarray}
F_x&\approx&\frac{g\mu_s\mu_0}{\pi\sqrt{2bW}\left(1-\gamma \sqrt{2\tilde{\lambda}}\right)}
\left\{1-
\tanh{\left[\frac{1}{2}\sqrt{\frac{\lambda}{b}}\right]}\right\},\\
F_z&\approx& -\frac{g\mu_s\mu_0}{\pi\sqrt{2bW}\left(1-\gamma \sqrt{2\tilde{\lambda}}\right)}
\left[1-\frac{\gamma}{\sqrt{2}} \frac{\left(\lambda/b\right)}{\sqrt{1+\left(\frac{\lambda}{5b}\right)^{2}}}\right].
\end{eqnarray}
\end{subequations}
\end{itemize}
\end{widetext}

Note how $\bm{F}_i$ close to the wire edge is sensitive to the ratio $\lambda/b$.  
The mid-edge case, Eq.~(\ref{fmax}) provides the maximum $|\bm{F}_i |$ for a single electron spin interacting with a thin-film wire. 
When $\lambda/b\lesssim 1$, 
$|\bm{F}_{{\rm edge}}|\approx \left(\frac{3.6\mu\textrm{m}}{\sqrt{bW}}\right)n\Phi_0$. However, when $\lambda/b\gg 1$ (ultra-thin wires),
the edge flux can be reduced by as much as a factor of three. 

At the edge, the associated local field produced by the SQUID's current on the spin is at most $|\bm{B}_{{\rm loc}}|=\mu_0I_{{\rm SC}}/(\pi \sqrt{bW})$. For typical $I_{{\rm SC}}\sim 1$~mA and $\sqrt{bW}\sim 1\;\mu$m we get $|\bm{B}_{{\rm loc}}|\sim 4$~Gauss, that is not sufficient to polarize electron spins even at the lowest temperatures achieved in the laboratory (10~mK). 
In SQUIDs made of carbon nanotubes or other nanostructures, the value of single-spin flux and the local field can be much larger.\cite{bouchiat09} 

\section{Numerical results and comparison to finite loop approximation/FastHenry\label{sec:numerres}}

We now present explicit numerical calculations of the flux vector by
numerical integration of Eqs.~(\ref{fxexp})--(\ref{fzexp}). In order to validate our numerical dipole method, we performed comparison calculations
using the finite loop/FastHenry method of
Refs.~\onlinecite{koch07},~\onlinecite{anton13b}.  We did this by designing a
FastHenry\cite{smithhisler96} input file that included a long SC wire
representing the SQUID, and a small square loop of side $0.1 \;\mu$m
representing the spin. FastHenry has the advantage of integrating the
London equations exactly.

Figure~\ref{fig:compare}(a) shows the computed $\Phi_X\equiv F_x/2$
(value of flux for spin-$1/2$ pointing along $x$) for electron spins interacting with a SC wire 
of penetration depth $\lambda=0.07\;\mu$m, thickness $b=0.1\;\mu$m, and lateral width $W=5.2\;\mu$m. The flux is
plotted as a function of spin location $X$ (along the lateral width as in
Fig.~\ref{Figcoord}). Each curve was calculated for a
different spin-wire surface distance $Z$. Note how the numerical
dipole calculations are right on top of the FastHenry results for 
$Z\leq 0.25$~$\mu$m. However, for $Z=0$ the numerical
dipole results are 50\% larger. Figure~\ref{fig:compare}(b) shows the
results for $\Phi_z\equiv F_z/2$ (flux for spin-$1/2$ pointing along
$z$, perpendicular to the wire surface). Again, we see that both
calculation methods agree for $Z\leq 0.25$~$\mu$m. However, at $Z=0$
(wire surface) we find that the numerical dipole method gives a flux
that is \emph{six times larger} than FastHenry.

Figure~\ref{fig:diponly} shows the numerical dipole results for
electron spins inside the wire (negative $Z$). While the flux is quite
high at the wire surface, it decreases to zero inside the wire
($Z=-b/2$ and $X$ away from the edges). This result is particularly
relevant for nuclear spins, as it shows that nuclei inside the
wire give a smaller contribution to flux noise (single nuclear spin flux is $\sim 10^{3}$ times smaller than the
single electron values shown in the figure). 

\begin{figure*}
\centering
\subfigure[Top surface, spin at $(X,0,0)$ or at $(X,0,-b)$]{\includegraphics[width=0.49\textwidth]{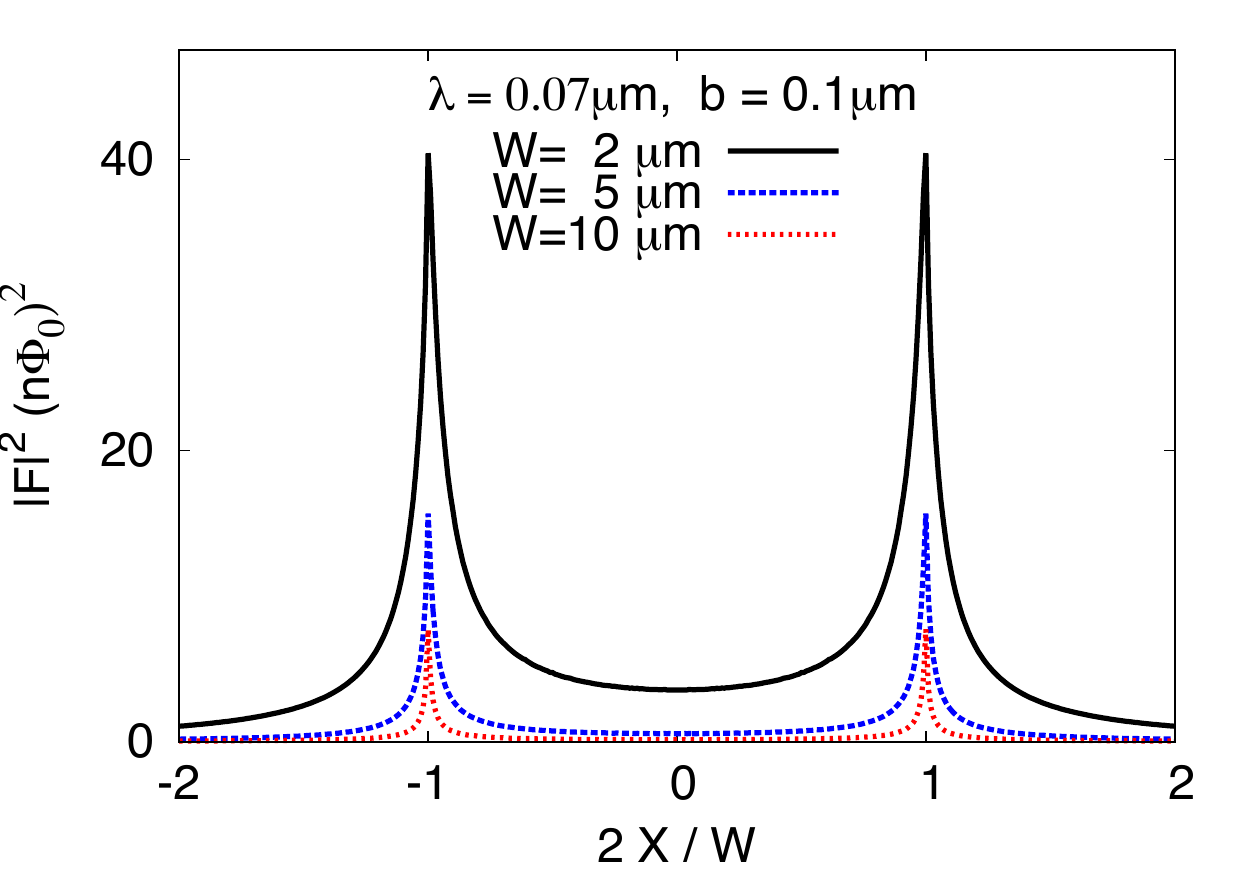} }
\subfigure[Edge surface, spin at $(\pm W/2,0,Z)$]{\includegraphics[width=0.49\textwidth]{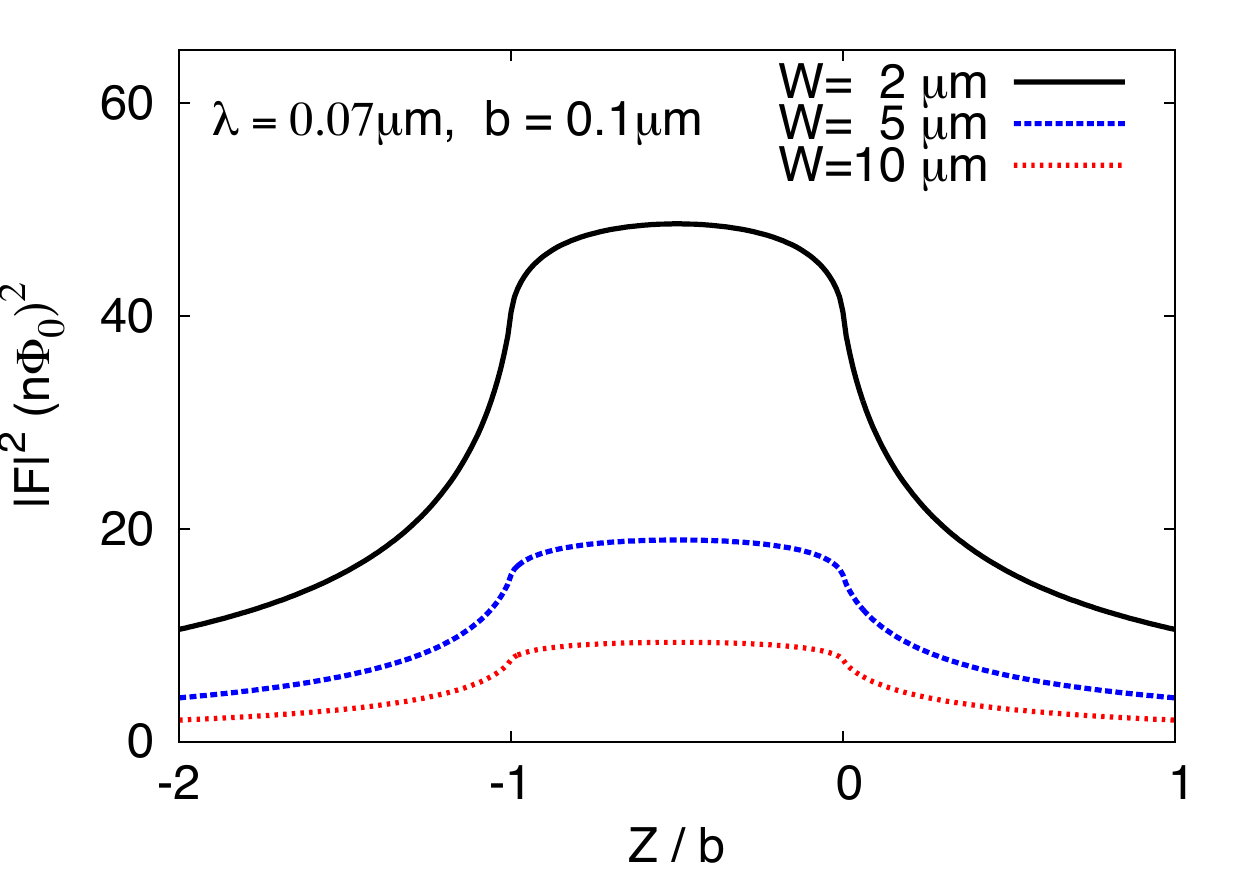}  }
\caption{(color online). Flux vector squared for electron spin at the top and edge surfaces of the wire, for $\lambda=0.07\;\mu$m, $b=0.1$~$\mu$m and $W=2, 5, 10$~$\mu$m. The coordinate $X$ runs along the lateral width of the top surface, while $Z$ runs along the side surface. Note how the flux is sharply peaked near the wire edges.\label{fig:fsquared}}
\end{figure*}

\begin{figure*}
\centering
\subfigure[Top surface, spin at $(X,0,0)$ or at $(X,0,-b)$]{\includegraphics[width=0.49\textwidth]{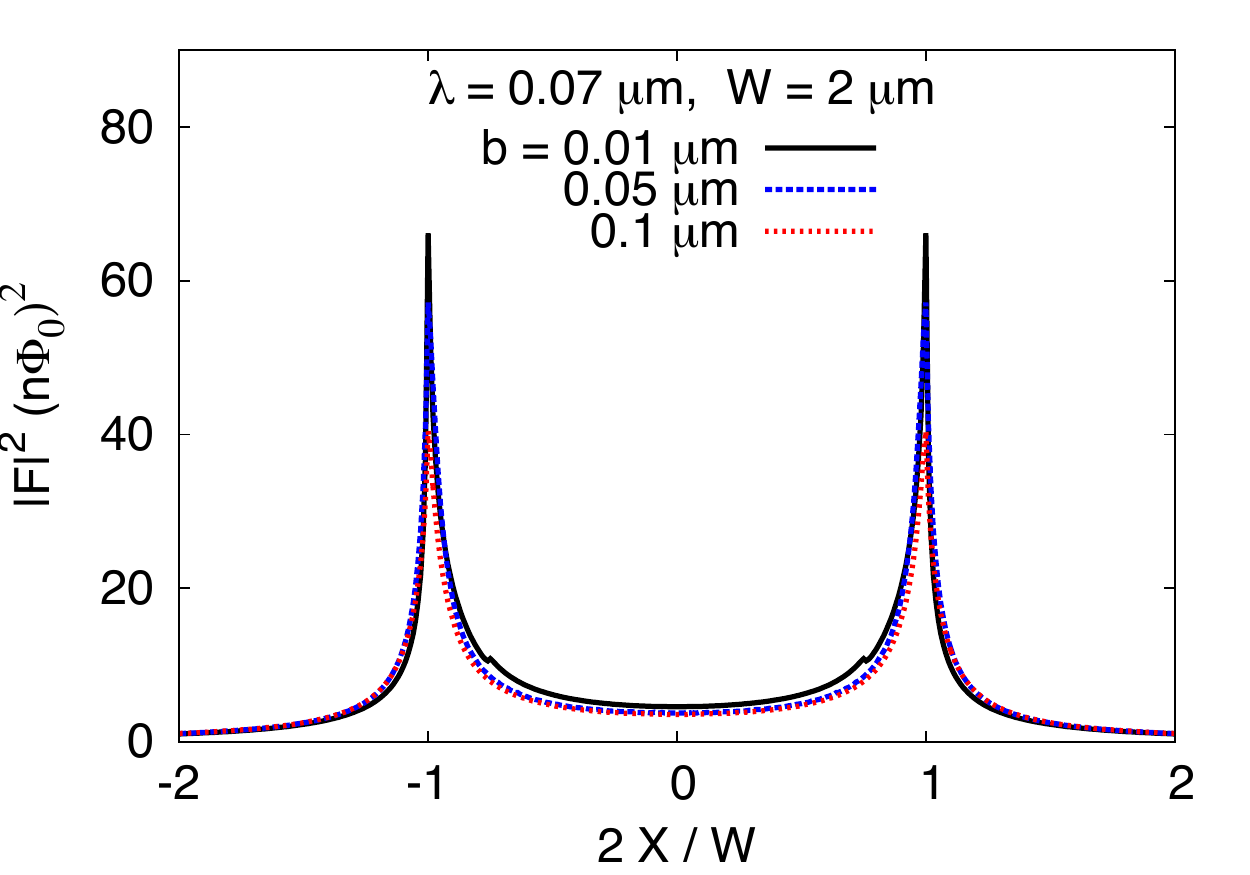}}
\subfigure[Edge surface, spin at $(\pm W/2,0,Z)$]{\includegraphics[width=0.49\textwidth]{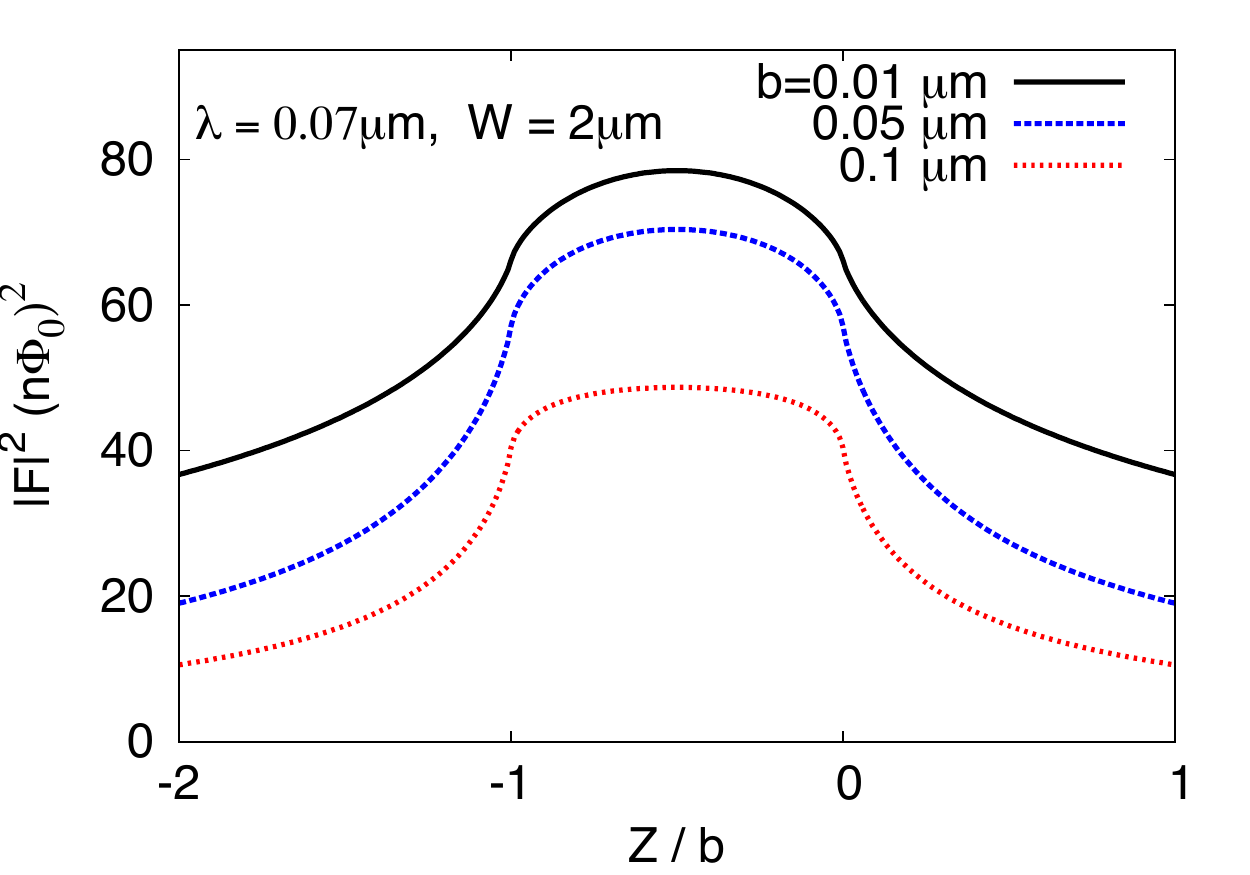}  }
\caption{(color online). Flux vector squared for electron spin at the top and edge surfaces of the wire, for $\lambda=0.07\;\mu$m, $W=2\;\mu$m, and  different wire thickness $b$. 
Note how the edge flux is quite sensitive to $b$.}
\label{fig:fsquaredbdepend}
\end{figure*}

\begin{figure*}
\centering
\subfigure[Top surface, spin at $(X,0,0)$ or at $(X,0,-b)$]{\includegraphics[width=0.49\textwidth]{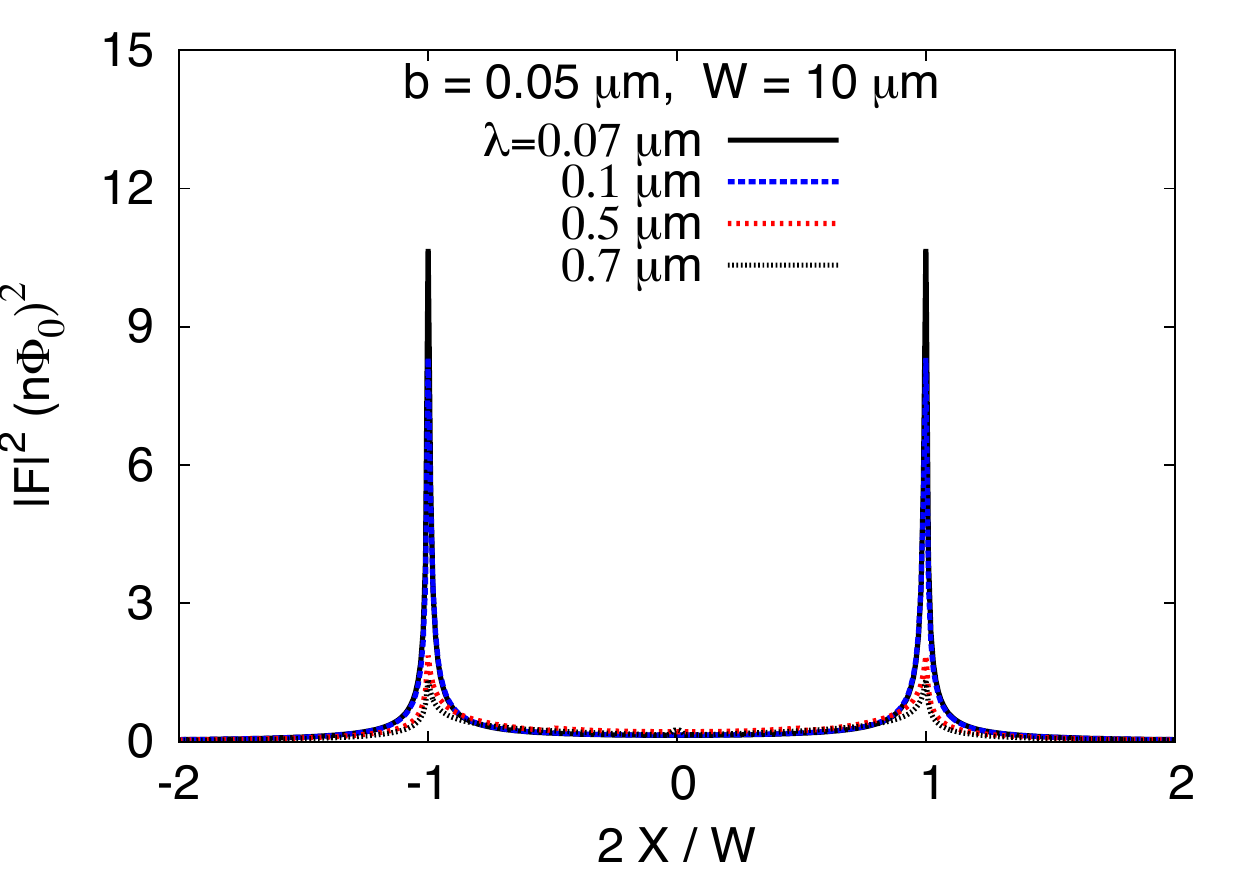} }
\subfigure[Edge surface, spin at $(\pm W/2,0,Z)$]{\includegraphics[width=0.49\textwidth]{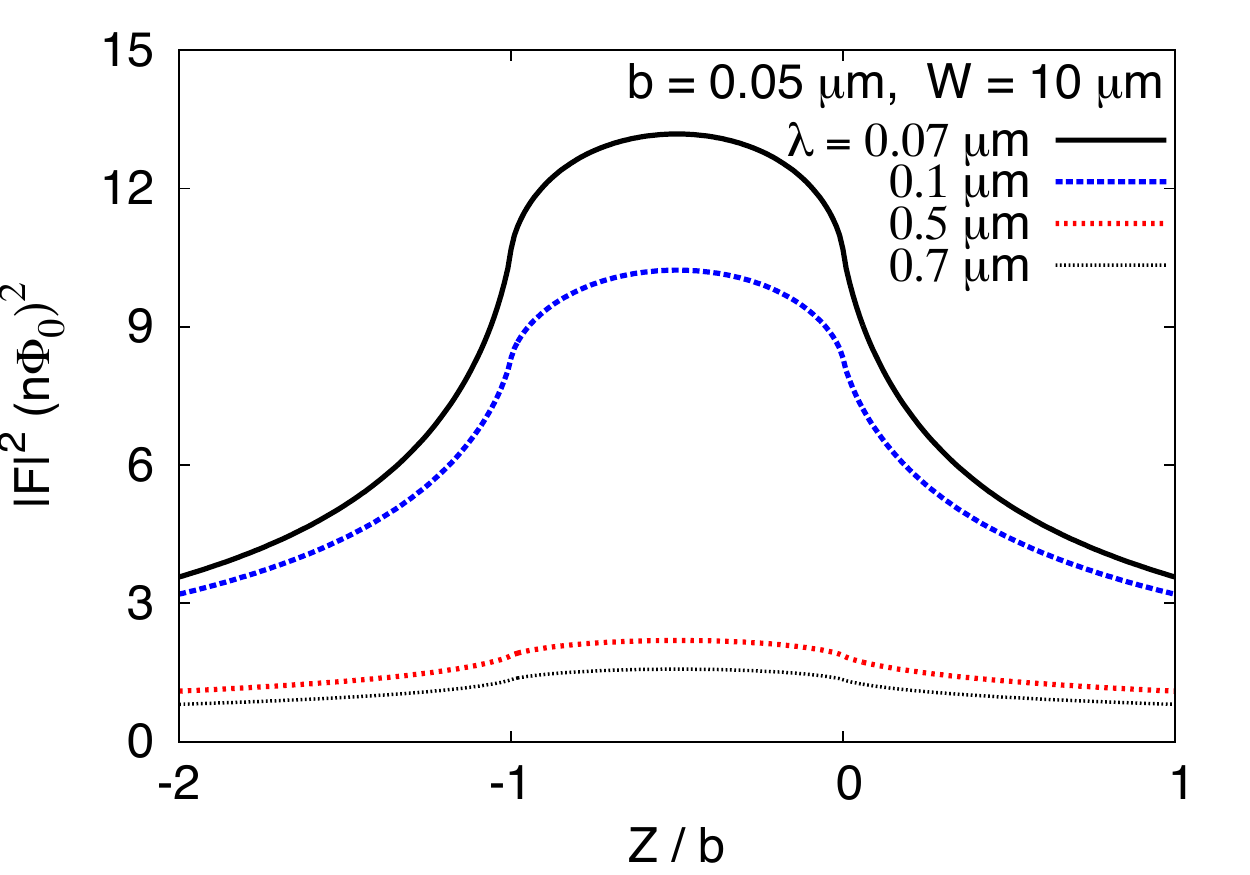}  }
\caption{(color online). Flux vector squared for electron spin at the top and edge surfaces of the wire, for $W=10\;\mu$m, $b=0.05\;\mu$m, and different penetration depth $\lambda$. Note how the edge flux is quite sensitive to $\lambda$.}
\label{fig:fsquaredlambdadepend}
\end{figure*}

It is important to note that all calculations presented here assumed a
SC wire of infinite length. As a result, the flux at the wire edges
(at $x=\pm W/2$) are of identical magnitude, i.e., the flux is
symmetric with respect to $X=0$. Actual devices will show some degree of asymmetry for the current densities at the wire edges. For example, the SQUID is a closed SC
wire loop, so to satisfy the Meissner effect it produces higher current 
density at the inside wire edge,\cite{anton13b} thus minimizing the value of magnetic field inside the wire. 
Modifying Eq.~(\ref{jsquid}) to include this asymmetry 
would lead to the same degree of asymmetry in the calculated flux vectors. 

Another source of asymmetry in $\bm{J}_{{\rm SC}}$ occurs due to spin polarization in applied $B$ fields. With $\bm{B}$ pointing along $\bm{\hat{x}}$, the superconductor will generate a current difference $\delta I_{{\rm SC}}$ between the top and bottom wire surfaces, in order to screen out the magnetic moment generated by polarized spins. A simple estimate is given by $|\delta I_{{\rm SC}}|L b= |\mu_B| N$, with $N\approx 2\sigma_2 LW$ the number of spins at the top+bottom surfaces.  This leads to $|\delta I_{{\rm SC}}|\sim 2\mu_B \sigma_2 W/b\sim 10-1000$~$\mu$A for $b=0.1\;\mu$m and $W=1-100\;\mu$m. 

While these asymmetries can be significant, they do not seem to modify the noise power results shown in the next section. This occurs because $\langle (\delta\Phi)^2\rangle$ is an integral of $|\bm{F}|^2$ over all wire surfaces; 
since the asymmetry increases the current in one region and decreases it by the same amount in another, the asymmetry cancels out in computations of the noise power summed over all surfaces. 

\section{Flux noise power for surface spins\label{sec:noisepowersurface}}

We will now present results for flux noise power in
a single wire with length $L$, width $W$, and thickness $b$. Our results can be applied to devices that contain more wires by
simply adding the noise power contributed by each wire segment.

\subsection{Without spatial correlation, no spin clusters\label{subsec:independent}}

Here we focus our discussion in the $T=\infty$ noise power with $S=1/2$ [Eq.~(\ref{noisepowerinfT})].
We recall that this constitutes the exact noise power for $S=1/2$ in the absence of spatial correlation and space inhomogeneity (See section~\ref{subsec:theorywithoutcorr}); 
and for $S>1/2$ it can be multiplied by $4S(S+1)$ to yield an upper bound on the uncorrelated noise at all temperatures 
[Eq.~(\ref{noisepoweruncorrineq})]. 

Figures~\ref{fig:fsquared}-\ref{fig:fsquaredlambdadepend} show explicit calculations of $|\bm{F}|^{2}$ for electrons located at the top and edge wire surfaces for a range of parameters. Here we see the extent to which
flux noise power is sharply peaked at $X=\pm W/2$ for all wire geometries. 

The noise power due to surface electrons in a single wire with length $L$ is evaluated 
by assuming a uniform area density of electrons $\sigma_2$ and plugging $S=1/2$ in Eq.~(\ref{noisepowerinfT}):
\begin{eqnarray}
\langle \Phi^{2}\rangle_{{\rm Top~Surf.}}&=& \frac{1}{4} \sigma_2 L \int_{-W/2}^{W/2} dX |\bm{F}|^{2}\nonumber\\
&\approx&  \frac{\frac{5}{16\pi^2}\left( g\mu_s\mu_0 \right)^{2}\left(\frac{W}{2b}\right)^{1/4}\left(\frac{\sigma_2 L}{W}\right)}{\left(1-\gamma\sqrt{2\tilde{\lambda}}\right)^2\left[1+2 \left(\frac{2b}{W}\right)^{2}\right]^{1/4}}\nonumber\\
&=& \frac{4.05  \left(\frac{W}{2b}\right)^{1/4}\left[\frac{\sigma_2 L (\mu{\rm m})}{W}\right](n\Phi_0)^{2}}{\left(1-\gamma\sqrt{2\tilde{\lambda}}\right)^2\left[1+2 \left(\frac{2b}{W}\right)^{2}\right]^{1/4}},
\label{powertop}
\end{eqnarray}
where in the second line we approximated the exact result by an analytic expression (good within 1\% for $10^{-3}<2b/W<1$), and in the third line we assumed $g\mu_s=2\mu_B$. 
We find that the top surface noise power is to a good approximation independent of $\lambda$.
The reason for this can be seen in
Fig.~\ref{fig:fsquaredlambdadepend}(a), where it is shown that $|\bm{F}|^2$ only depends on $\lambda$ when $X$ is extremely close to $\pm W/2$ (Note how the peak value increases with decreasing $\lambda$, but the peak width remains small). As a result we find that the $\lambda$-dependent contribution to Eq.~(\ref{powertop}) is always small.

The edge contribution is quite different. Figures~\ref{fig:fsquared}(b)-\ref{fig:fsquaredlambdadepend}(b) show that it is a good approximation to assume $|\bm{F}|^2$ constant for all $Z/b$, 
with its value given by Eq.~(\ref{fmax}) squared. Hence we get: 
\begin{eqnarray}
\langle \Phi^{2}\rangle_{{\rm Edge~Surf.}}&\approx& \frac{1}{4} \sigma_2 L b |\bm{F}_{{\rm edge}}|^{2}\nonumber\\
&=& 
\frac{\left( g\mu_s\mu_0 \right)^{2}}{4\pi^2}\left(\frac{\sigma_2 L }{W}\right)\left(\frac{1-\gamma \frac{\lambda/b}{\sqrt{1+2\left(\frac{\lambda}{5b}\right)^{2}}}}{1-\gamma\sqrt{2\tilde{\lambda}}}\right)^{2}\nonumber\\
&=& 3.24\left[\frac{\sigma_2 L (\mu{\rm m})}{W}\right]\left(\frac{1-\gamma \frac{\lambda/b}{\sqrt{1+2\left(\frac{\lambda}{5b}\right)^{2}}}}{1-\gamma\sqrt{2\tilde{\lambda}}}\right)^{2}\nonumber\\
&&\times (n\Phi_0)^{2}.
\label{powerside}
\end{eqnarray}
Comparing Eq.~(\ref{powerside}) to Eq.~(\ref{powertop}) we see that in
a typical device (satisfying $W<10^3 b$), the edge surface
contribution has the same order of magnitude as the top surface
contribution.  Flux noise measurements in several niobium
SQUIDs\cite{lanting14} were fitted to spin-diffusion theory to yield $\sigma_2=5\times 10^{12}$~cm$^{-2}$, leading to noise power 
Eqs.~(\ref{powertop})~and~(\ref{powerside}) in the range of $1-100$~$(\mu\Phi_0)^{2}$,
see Fig.~\ref{fig:noisecomp}.

Our expressions for the noise power account for the spin-dipole singularity fully, giving values that are qualitatively different and numerically much larger than other expressions derived in the literature. 
For example, Eq.~(6) of Ref.~\onlinecite{bialczak07} predicted a term proportional to $\ln{(\tilde{\lambda})}$ in the top surface electron noise power of a circular SQUID; in contrast,
our results show that the top noise power is roughly independent of $\tilde{\lambda}$, with the edge noise strongly dependent on $\lambda/b$. 

We now compare our results to the state of the art numerical
calculations of noise power in SQUID washers presented in Anton {\it
  et al.}\cite{anton13b} Our Eq.~(\ref{noisepowerinfT}) is larger than
Eq.~(1) of Ref.~\onlinecite{anton13b} by a factor of $S(S+1)/S^{2}$,
which equals $3$ for $S=1/2$.  After multiplying the results of
Ref.~\onlinecite{anton13b} by $3$, we find that our
$\left\langle\Phi^2\right\rangle_{{\rm Top + bottom}}$ (obtained after
adding-up $4$ wire segments forming a washer) is 10\% smaller than
theirs for the case of short wires ($L\sim 10\;\mu$m) in a wide range
of parameters.  For longer wires ($L\sim 100\;\mu$m), we find that our
results are as much as 40\% smaller. This quite good agreement
indicates that corner effects are not substantial.

\subsection{With spatial correlation,  ferromagnetic spin clusters\label{subsec:corr}}

We now present explicit numerical results of noise power with
ferromagnetic correlations, for the case of surface electrons. In
order to evaluate Eq.~(\ref{corrnoise2d}) we separate each surface
integral into the $4$ surfaces making the wire:
\begin{equation}
\int d^{2}r = \int_{{\rm Top}} d^{2}r + \int_{{\rm Bottom}} d^{2}r + \int_{{\rm EdgeR}} d^{2}r + \int_{{\rm EdgeL}} d^{2}r. 
\end{equation}
The total integral $\int d^{2}r\int d^{2}r'$ breaks up into $16$ contributions, of
which only $5$ are distinct: $2\times$Top-Top (because Bottom-Bottom
is identical to Top-Top by symmetry), $2\times$EdgeR-EdgeR,
$8\times$Top-Edge, $2\times$Top-Bottom, and $2\times$EdgeL-EdgeR.  For
ferromagnetic correlations the first $3$ are always positive, while
the last two are always negative.

\begin{figure}
  \centering
  \includegraphics[width=0.49\textwidth]{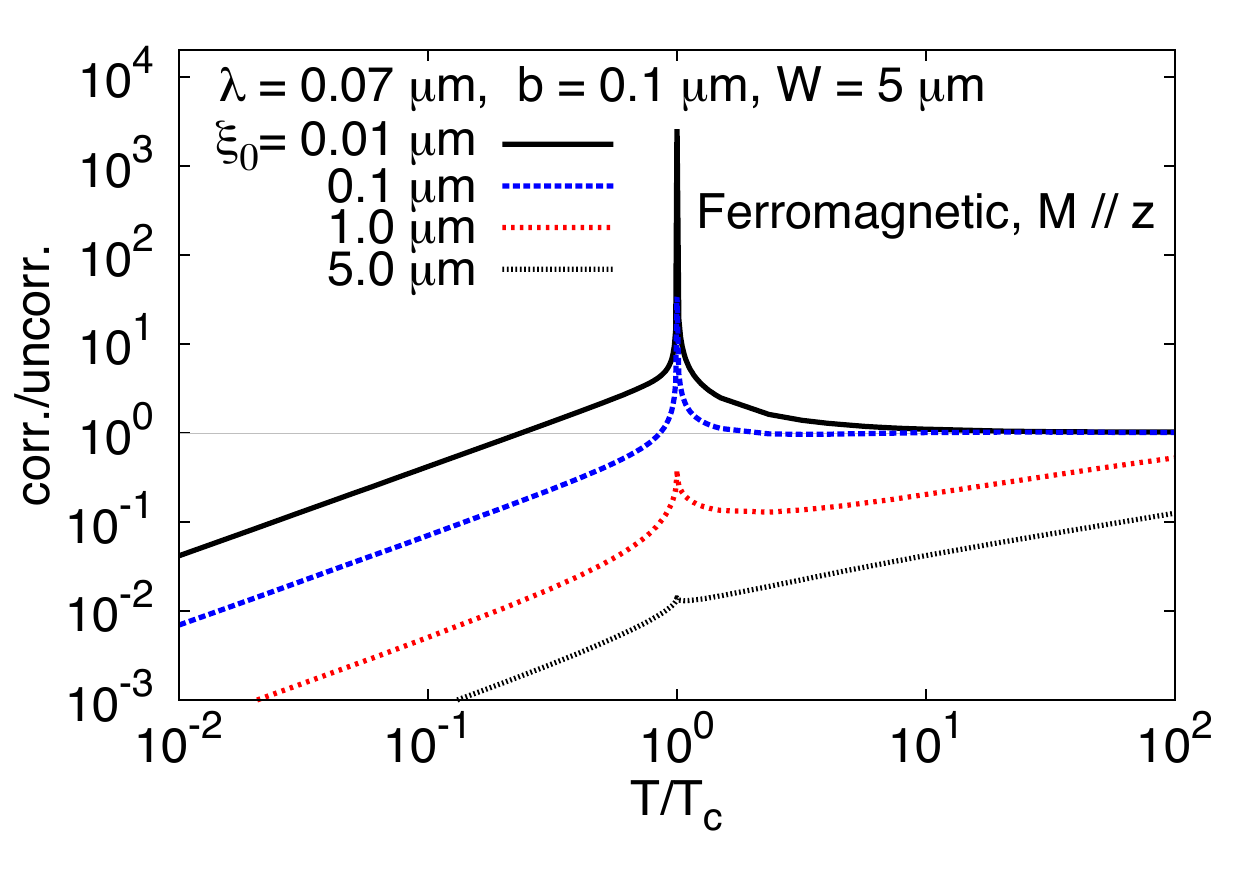} 
  \caption{(color online). Effect of a ferromagnetic phase transition
    on flux noise power, calculated using mean-field theory with anisotropy energy
 $K/(\sigma_2\xi_{0}^{2}\bar{J})=0.1$ along easy axis $\bm{\hat{e}}_{\parallel}=\bm{\hat{z}}$.  The ratio
    of correlated noise to $T=\infty$ noise is shown as a function of
    $T/T_c$, calculated from Eq.~(\ref{corrnoise2d}). At $T\approx
    T_c$ a sharp finite peak appears, with linewidth and peak value
    depending on $\xi_0$, the range of the exchange interaction.  For
    large $\xi_0$, correlated noise can be several orders of magnitude
    smaller than the $T=\infty$ noise in the region $T_c\ll T<\infty$, 
    suggesting a new method to reduce flux noise in SC
    devices. The noise reduction occurs due to ferromagnetic correlation between the top and bottom surfaces with antiparallel $\bm{F}$.}
\label{fig:CorrNoiseTcCompxi0}
\end{figure}

Figure~\ref{fig:CorrNoiseTcCompxi0} shows calculations of the
correlated noise power Eq.~(\ref{corrnoise2d}) divided by the $T=\infty$ uncorrelated noise Eq.~(\ref{noisepowerinfT}) (this ratio is denoted corr./uncorr.),
as a function of $T/T_c$, for anisotropy energy $K/(\sigma_2\xi_{0}^{2}\bar{J})=0.1$ along the easy-axis $\bm{\hat{e}}_{\parallel}=\bm{\hat{z}}$ (as $T\rightarrow 0$ all spins will polarize along $\bm{\hat{e}}_{\parallel}$). 
The correlated noise power has a sharp peak at $T=T_c$; within
mean-field theory this peak is finite, and has its width and height
governed by the exchange interaction range $\xi_0$. Quite
interestingly, for $\xi_0\gg b$ the correlated noise can be several orders of magnitude smaller than
uncorrelated noise in all regimes, including the $T_c\ll T\ll
\infty$ regime where mean-field theory is known to be accurate.

\begin{figure*}
  \centering \subfigure[Total ratio for different
  $W$\label{fig:CorrNoiseWcomp}]{\includegraphics[width=0.49\textwidth]{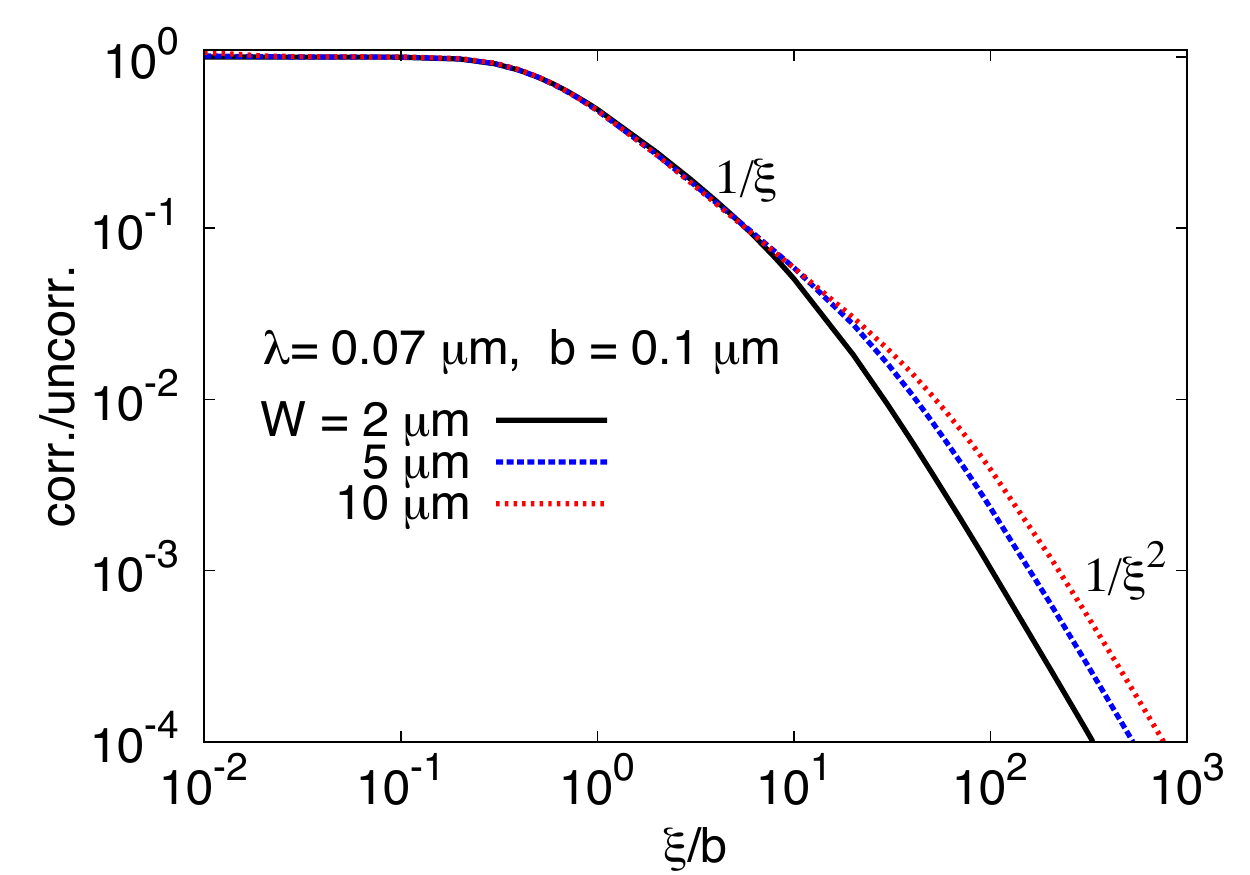}}
  \subfigure[Individual contributions for $W=5\;\mu$m\label{fig:CorrNoiseWcompItems}]{\includegraphics[width=0.49\textwidth]{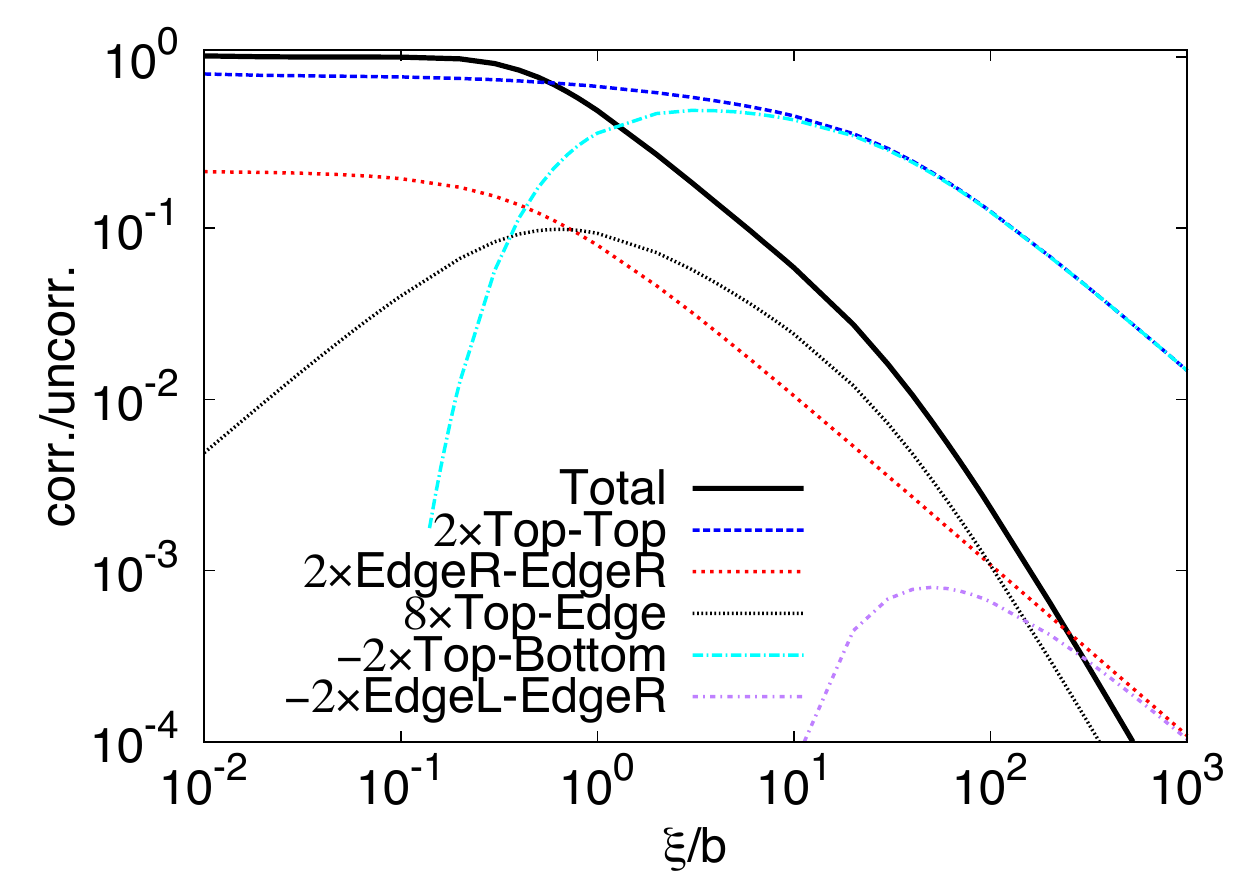}}
\caption{(color online). Noise power for the ferromagnetic model with $\bm{\hat{e}}_{\parallel}=\bm{\hat{z}}$ (same as Fig.~\ref{fig:CorrNoiseTcCompxi0}) in the $T\gg T_c$ regime.
(a) Correlated noise power divided by
  uncorrelated noise power (at $T=\infty$). The results are
  presented as a function of average spin cluster size $\xi_{\parallel}\approx\xi_{\perp}\equiv\xi$ divided
  by wire thickness $b$. Note how the presence of ferromagnetic spin
  clusters makes the correlated/uncorrelated ratio smaller than one in
  the $T_c\ll T<\infty$ regime. The origin of this effect is explained in
  (b), where each contribution is labelled by a pair of surfaces; for
  example, $-2\times$Top-Bottom refers to the negative of Top-Bottom +
  Bottom-Top contribution in Eq.~(\ref{corrnoise2d}).  When $\xi>b$, ferromagnetic
  spin-spin correlation between the wire's top and bottom surfaces
  (where $\bm{F}$ is antiparallel) is activated, reducing correlated
  noise power to a value below the $T=\infty$ uncorrelated noise. When
  $\xi>W$, ferromagnetic correlation between wire corners (where $\bm{F}$ is also
  antiparallel) is activated, producing further reduction of
  correlated noise.\label{fig:CorrNoiseDependxi}}
\end{figure*}

To shed light in this feature, Fig.~\ref{fig:CorrNoiseWcomp} shows correlated noise for $T\gg T_c$, as a function of $\xi_{\parallel}\approx
\xi_{\perp}\equiv \xi$.  Here we see that the origin of the
noise reduction is the presence of spin clusters (non-zero spatial correlation).  \emph{Remarkably, correlated noise is
  always smaller than uncorrelated noise for all average cluster sizes
  $\xi$}. Figure~\ref{fig:CorrNoiseWcompItems} explains the origin of
the reduction: For $\xi>b$, the negative Top-Bottom contribution (with
$\bm{F}$ antiparallel) is activated, which nearly cancels out the
Top-Top contibution when $\xi\gg b$. Thus, 
antiparallel $\bm{F}$ for surfaces across from each other make inter-surface ferromagnetic fluctuations interfere \emph{destructively} with the intra-surface ferromagnetic fluctuations, leading to a reduction of flux noise.

While uncorrelated noise scales roughly as $\sigma_2
L/W$,\cite{koch07,bialczak07} correlated noise shows different
behavior governed by the additional length scales $\xi$. In the
``short-range order'' regime $b\ll\xi\ll W$, correlated noise scales
instead as $\sigma_2 L\lambda/(\xi b)$; in the ``long-range order''
regime with $\xi \gg W$ it scales as $\sigma_2 L\lambda W/(b\xi^{2})$. 

When $T\ll T_c$ the spins are polarized along $\bm{\hat{e}}_{\parallel}$ leading to pure quantum fluctuation noise that is larger than the mean-field theory prediction.  
At $T=0$ the noise can be calculated exactly using Eq.~(\ref{powerBlimit}) for  $\bm{\hat{B}}=\bm{\hat{e}}_{\parallel}$ leading to the values calculated in Table~\ref{tableepar}. 

\begin{table} 
\begin{center} 
\begin{tabular}{c |c } 
$\bm{\hat{e}}_{\parallel}$ & corr./uncorr. at $T=0$\\
\hline 
$\bm{\hat{x}}$ & $0.44/(S+1)$\\
$\bm{\hat{y}}$ & $1.5/(S+1)$\\
$\bm{\hat{z}}$ & $1.1/(S+1)$\\
\end{tabular} 
\caption{Exact flux noise [from Eq.~(\ref{powerBlimit})] for the spin-polarized ground state for three spin polarization directions $\bm{\hat{e}}_{\parallel}$. We assumed 
a long wire with $\bm{J}_{SC}\parallel \bm{\hat{y}}$ and other coordinates as in Fig.~\ref{Figcoord}. The zero point flux fluctuation goes to zero only in the classical limit $S\rightarrow \infty$.}
\label{tableepar} 
\end{center} 
\end{table} 

\begin{figure}
  \centering
  \includegraphics[width=0.49\textwidth]{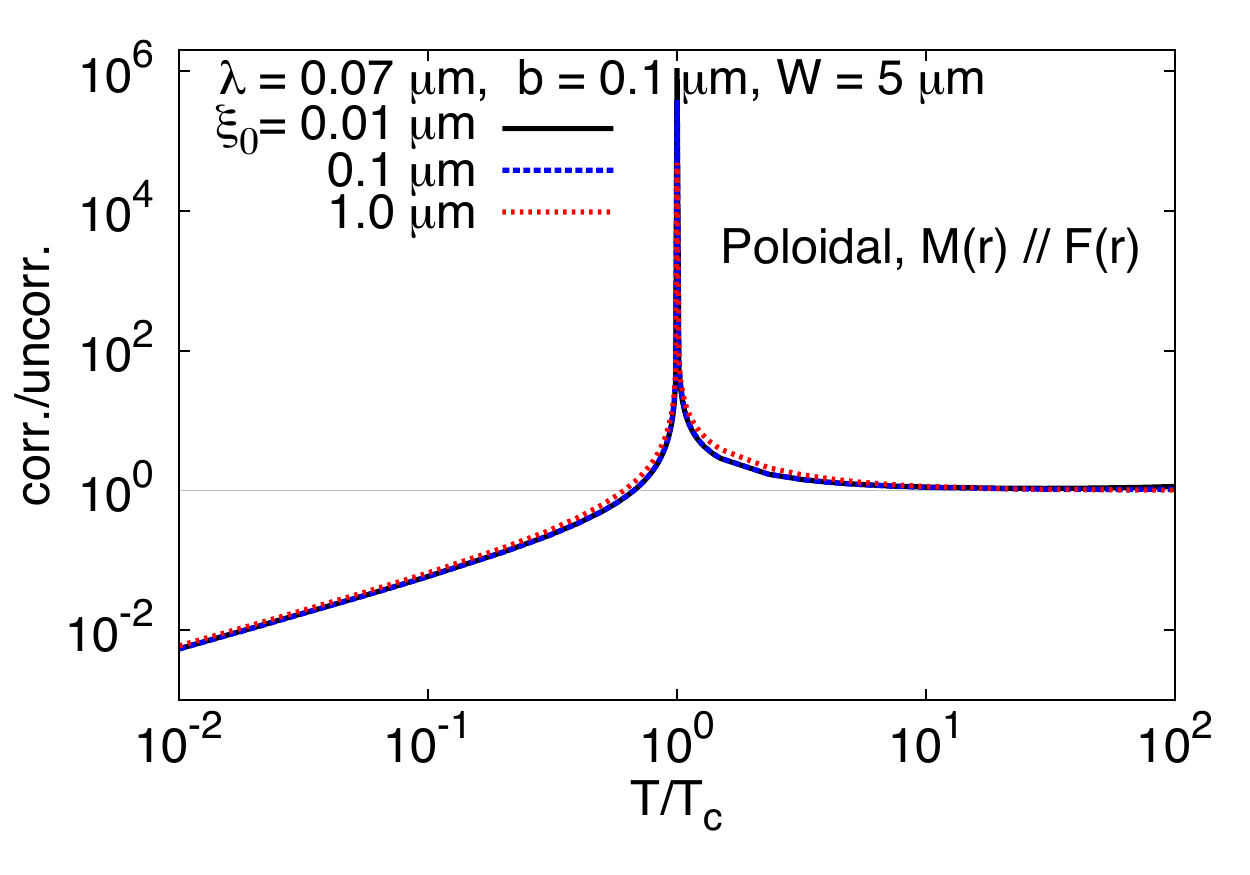} 
  \caption{(color online). Flux noise power of the poloidal model [Eq.~(\ref{hexpanded}) with $\bm{\hat{e}}_{\parallel}(\bm{r})=\bm{\hat{F}}(\bm{r})$] with 
  anisotropy energy $K/(\sigma_2\xi_{0}^{2}\bar{J})=100$.
  The ratio of correlated noise to $T=\infty$ noise is shown as a function of $T/T_c$, calculated from Eq.~(\ref{corrnoise2d}). At $T>T_c$ flux noise is always larger than its $T=\infty$ value, in contrast to the result shown for the ferromagnetic phase  (Fig.~\ref{fig:CorrNoiseTcCompxi0}). This occurs because the poloidal model has antiferromagnetic correlation between the top and bottom surfaces. At $T=0$ the poloidal state has zero flux noise  
 [Eq.~(\ref{fnpoloidal})] similar to the result obtained in mean-field theory. Engineering the spin system to be close to the poloidal phase provides a novel method to reduce flux noise in SC devices.}
\label{fig:CorrNoiseTcCompxi0Poloidal}
\end{figure}

As noted in Eq.~(\ref{powerBlimit}), flux noise is exactly zero for the poloidal state. 
We now describe the flux noise power as a function of $T$ for the poloidal model, which comprises 
Eq.~(\ref{hexpanded}) with $\bm{\hat{e}}_{\parallel}(\bm{r})=\bm{\hat{F}}(\bm{r})$ [as noted above this model has the poloidal state as its ground state when $K\gg \bar{J}\sigma_d \xi_{0}^{d} (\xi_0/b)^{2}$]. The resulting correlation functions are formally identical to Eqs.~(\ref{cpar2d})~and~(\ref{cperp2d}); however, they describe correlations between $\bm{\hat{e}}_{\parallel}$ directions that are different in different surfaces (we take $\bm{\hat{e}}_{\perp 1}=\bm{\hat{y}}$ and $\bm{\hat{e}}_{\perp 2}=\bm{\hat{F}}(\bm{r})\times\bm{\hat{y}}$). For example, $M_x$ in the top surface is correlated to $-M_z$ in the edge right surface, and to $-M_x$ in the bottom surface.  As a consequence, the poloidal model shows antiferromagnetic correlation between spins located in surfaces across from each other, with ferromagnetic correlations between spins located in the same surface. The resulting flux noise power as a function of $T/T_c$ is shown in Fig.~\ref{fig:CorrNoiseTcCompxi0Poloidal}. The behavior of the poloidal model is seen to be quite distinct from the case of pure ferromagnetic correlation. 
The noise at $T\approx T_c$ is more than $100$ times larger, and at $T>T_c$ correlated flux noise power is \emph{always larger} than the $T=\infty$ uncorrelated case. Clearly, this occurs because for $\xi>b$
the antiferromagnetic inter-surface correlated fluctuations interfere \emph{constructively} with the intra-surface ferromagnetic fluctuations; a simple consequence of having antiparallel $\bm{F}$ between surfaces across from each other.
In contrast to the behavior shown in Figs.~\ref{fig:CorrNoiseDependxi}(a,b) noise power \emph{increases} with increasing spin correlation length $\xi$. When $T>T_c$ and $b\ll\xi\ll W$, correlated noise scales
as $\sigma_2 L\lambda \xi/b^{3}$; in the ``long-range order''
regime with $\xi \gg W$ it saturates at its maximum value proportional to $\sigma_2 L \lambda/b^2$.

\section{Bulk nuclear spins\label{sec:noisepowerbulk}}

For bulk lattice spins inside the wire Eq.~(\ref{noisepowerinfT}) becomes
\begin{equation}
\langle \Phi^{2}\rangle_{{\rm Bulk}}=\frac{S(S+1)}{3}\sigma_3 \int_{{\rm wire}} d^3 r\left|\bm{F}(\bm{r})\right|^2, 
\end{equation}
with the integral running over the volume of the wire, and $\sigma_3$ the corresponding volume density for spins. We evaluated this expression numerically and were able to fit the following expression with high accuracy in the region $10^{-3}<2b/W<1$:
\begin{equation}
\langle \Phi^{2}\rangle_{{\rm Bulk}}\approx \frac{S(S+1)}{6\pi^4}\left( g\mu_s\mu_0 \right)^{2} \left(\frac{\sigma_3 LW}{b}\right) \frac{\ln{\left[1+\left(\frac{4b}{W}\right)^2\right]}}{\left(1-\gamma\sqrt{2\tilde{\lambda}}\right)^2}.
\label{powerbulk}
\end{equation}
Like the case of top surface noise, Eq.~(\ref{powerbulk}) is independent of $\lambda$ because at the edge $\bm{F}^2$ depends on $\lambda$ only in a very small fraction of the wire volume. This dependence becomes negligible after volume integration.

We present numerical results for bulk nuclear spins in the typical superconductors aluminum and niobium. For aluminum, the $^{27}$Al isotope is 100\% abundant in nature, has $S=5/2$ 
and $g=1.46$,\cite{brown56} and forms a fcc lattice with lattice parameter $4.05$~\AA. Thus, $\sigma_3=4/(4.05\;{\rm \AA})^3=6.02\times 10^{10}\;\mu{\rm m}^{-3}$, and from Eq.~(\ref{powerbulk}) we get
\begin{equation}
\langle \Phi^{2}\rangle_{{\rm Al}} =1.8\times 10^{-2} \left[\frac{LW}{b(\mu m)}\right] \frac{\ln{\left[1+\left(\frac{4b}{W}\right)^2\right]}}{\left(1-\gamma\sqrt{2\tilde{\lambda}}\right)^2} (\mu\Phi_0)^2.
\end{equation}
For niobium, the $^{93}$Nb isotope is 100\% abundant in natural samples, with $S=9/2$ and $g=1.37$,\cite{sheriff51} and forms a bcc lattice with parameter $3.30$~\AA~leading to $\sigma_3=2/(3.30\;{\rm \AA})^3=5.56\times 10^{10}\;\mu{\rm m}^{-3}$. We get 
\begin{equation}
\langle \Phi^{2}\rangle_{{\rm Nb}} =4.2\times 10^{-2} \left[\frac{LW}{b(\mu m)}\right] \frac{\ln{\left[1+\left(\frac{4b}{W}\right)^2\right]}}{\left(1-\gamma\sqrt{2\tilde{\lambda}}\right)^2} (\mu\Phi_0)^2.
\label{powernb}
\end{equation}
The Nb noise power is 2.3 times larger than Al. 

\begin{figure}
  \centering
  \includegraphics[width=0.49\textwidth]{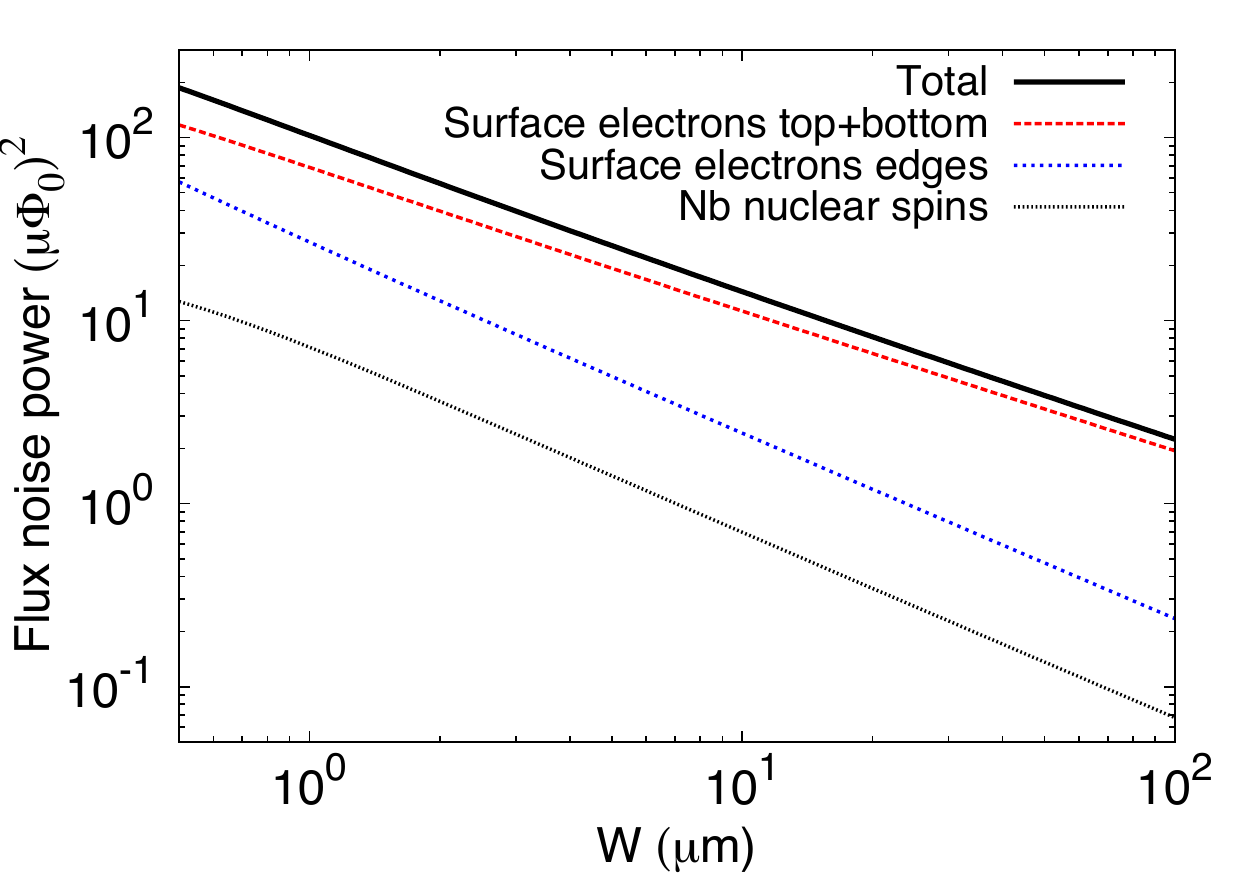} 
  \caption{(color online). Contributions to the $T=\infty$ flux noise
    power due to surface electrons and nuclear spins, for a niobium
    wire loop with length $L=100$~$\mu$m, penetration depth
    $\lambda=0.07\;\mu$m, and thickness $b=0.1$~$\mu$m, as a function
    of wire lateral width $W$. For the surface electrons, we assumed
    spin area density $\sigma_2=5\times 10^{12}$~cm$^{-2}$ (as
    measured in Ref.~\onlinecite{lanting14}), and separated the
    contributions into top+bottom surfaces and inner+outer side
    surfaces. The nuclear spin contirbution is for Nb; the noise for
    Al would be 2.3 times smaller.}
\label{fig:noisecomp}
\end{figure}

Figure~\ref{fig:noisecomp} compares the $T=\infty$ surface electron and nuclear
spin contributions for a Nb wire loop with $L=100\;\mu{\rm m}$,
$b=0.1\;\mu{\rm m}$, and $\sigma_2=5\times 10^{12}$~cm$^{-2}$. It
shows that the contribution of bulk nuclear spins to the flux noise
power is typically 5\% of the total noise, that is dominated by
surface electrons close to the wire edges.

\section{Discussion and conclusion\label{sec:conclusions}}

We proposed a flux-vector model of flux noise due to spins in
superconducting devices, and performed explicit numerical calculations
of the flux noise power produced by localized electron and nuclear
spins. We emphasized the crucial difference between electron impurities
and lattice nuclear spins.  Electron impurities are typically
concentrated in the wire surface (where flux vector $\bm{F}$ is
maximum), and can be substantially affected by the formation of spin
textures above or below the critical temperature for a spin phase transition. Nuclear spins are instead
distributed in the bulk of the wire, and their noise is well described
by the $T=\infty$ limit.

In many cases the vectorial nature of the spin-wire coupling plays an
essential role in determining the value of the noise power.  This
includes the case of a phase
transition, when additional correlation length scales appear in the
problem (describing the average size of spin clusters). Even in
the absence of a phase transition the vector nature of $\bm{F}$ plays
a role, e.g. for spins with easy-axis anisotropy (typical of $S>1/2$
transition metal impurities).  The key experiment to directly verify
the vectorial nature of $\bm{F}$ is to measure flux noise in SQUIDs as
a function of the magnitude and in-plane direction of an applied magnetic
field. Fields of $0.5$~Tesla at temperatures of $0.1$~K will polarize
electron spins without affecting nuclear spins.  
Table~\ref{tableepar} compares values of flux noise power for large $\bm{B}$ applied along different directions. Fits using Eq.~(\ref{powerB}) plus a field independent
contribution will allow measurement of $S$ and disentanglement of
electron and nuclear spin contributions. These are quite challenging
measurements, but we hope they will be performed in the future to
confirm the origin of flux noise in superconducting devices.

Our explicit calculations of $\bm{F}$ for thin-film superconducting
wires shows that the flux at the wire edges is much larger than
anticipated by previous calculations\cite{koch07,bialczak07,desousa07}
because of two reasons.  First, the singularity associated to the
spin-dipolar field enhances the flux at the wire edges; second, the
edge surfaces of thin-film wires (hitherto ignored in previous
calculations) contributes the same order of magnitude as the top +
bottom surfaces. As a result, the scaling relations for flux as a
function of wire geometry are modified.  Flux noise can be greatly reduced 
by minimizing the
electron spin density at the wire edge region. One might be able to
achieve this with chemical passivation of the
surface\cite{desousa07,lee14} or by growing the wires with layer by
layer deposition (to reduce the number of vacancies and other defects)
instead of the usual evaporation method.

We also presented for the first time a realistic estimate for the
noise power contributed by nuclear spins. Nuclear spin flux noise has
been a subject of speculation for several
years,\cite{sleator85,dube01,rose01,koch07,wu12} and we can now
ascertain that it accounts for approximately $5\%$ of the total flux
noise power affecting typical superconducting devices made with
aluminum or niobium. In the future, it is conceivable that one will be
able to design SQUIDs with much lower defect spin density, making
lattice nuclear spins the ultimate source of noise to be optimized.
Nuclear spin noise spectral density ranges from $0$~Hz up to
$\sim 10^{4}$~Hz, the value of nearest neighbor spin-spin dipolar
interaction.  For very thin wires ($2b/W\ll 1$), nuclear spin noise
will scale proportional to $bL/W$, so that further reduction can be
achieved by reducing $b,L$ and increasing $W$. Another alternative
would be to use materials with zero lattice nuclear spin, such as
making the superconducting wires with lead (Pb). Natural lead samples
have 77.9\% of zero-nuclear-spin isotopes ($^{204}$Pb, $^{206}$Pb, and
$^{208}$Pb), with nuclear spin $S=1/2$ present in only 22.1\%
($^{207}$Pb). Hence, a dramatic reduction in nuclear spin noise is
predicted for natural Pb wires.  Nuclear spin noise can also be
reduced in niobium by using metastable nuclear states such as
$^{93m}$Nb, the first excited state of $^{93}$Nb (100\% natural
abundance). $^{93m}$Nb has $S=1/2$ with a half-life of 16
years,\cite{baglin97} providing noise reduction by a factor of
$(\frac{1}{2}\times\frac{3}{2})/(\frac{9}{2}\times\frac{11}{2})=\frac{1}{33}$.

For electron spins, a critical experimental question is whether and
how flux noise power depends on temperature. The proper answer to this
question should reveal whether or not the spin system is close to a
phase transition.  So far two experiments addressed this question,
reaching opposite conclusions in different temperature ranges.
Ref.~\onlinecite{anton13} measured a change of up to $10^{3}$ in SQUID
flux noise power when the temperature changed from $0.1-4$~K; in
contrast, Ref.~\onlinecite{lanting14} made a direct measurement of
noise power in the lower temperature range $0.01 - 0.1$~K (see their
Fig.~3), and concluded that there was no change within experimental
error bars. It should be noted that even in the absence of a phase
transition the noise power may be temperature dependent. This can
occur because of spin anisotropy [Eq.~(\ref{noiseanis})] or because
$\bm{F}(\bm{r})$ is temperature-dependent at temperatures near
the superconducting critical temperature [when penetration depth
$\lambda$ increases with temperature and Eq.~(\ref{lambdatilde})
no longer holds]. The latter effect might be contributing to the
temperature dependence observed in Ref.~\onlinecite{anton13}.

Our results provide guidance on how to reduce flux noise in SQUIDs and
superconducting qubits by changing the wire geometry and/or
engineering a spin phase transition. Our Fig.~\ref{fig:noisecomp}
confirms the original claim\cite{bialczak07,koch07} that uncorrelated noise
scales roughly as $\sigma_2 L/W$, and can be greatly reduced by using
wider wires with larger $W$.  However, in the presence of spin-spin
correlation the geometrical scaling of noise is quite different due to
the introduction of the spin cluster length scales $\xi$.  

We make several predictions for flux noise power in the presence of magnetic correlation. If the spins at the top surface couple ferromagnetically (antiferromagnetically) to the spins at the bottom surface, flux noise power gets reduced (increased) when the correlation length scale $\xi$ is increased. We should mention that magnetic coupling across thin metallic surfaces is a well established phenomenon in the physics of 
normal metals sandwiched between ferromagnetic layers. For example, Ref.~\onlinecite{parkin91} demonstrates that the RKKY interaction can induce magnetic coupling that alternates back and forth between antiferromagnetic and ferromagnetic depending on the thickness and electron density of the normal metal layer. Perhaps in the future we will be able to engineer SC wires with similar alternation between ferro and antiferro couplings, allowing the corresponding control over flux noise induced by electron spin impurities at the wire surface.

Moreover, we predict that flux noise power is exactly equal to zero for the poloidal spin texture (all spins $\bm{s}_i$ polarized parallel or antiparallel to their corresponding flux vector $\bm{F}_i$). This occurs 
because at $T=0$ spin fluctuation is perpendicular to the spin polarization axis, which for the poloidal texture is perpendicular to $\bm{F}_i$ by design; as a result magnetic fluctuation of the poloidal state does not produce flux noise. How to imprint the poloidal texture to a disordered spin system is a question for future research. Given that $\bm{F}_i$ is parallel to the magnetic field generated by the SC wire, it might be possible to at least partially polarize the spins in the poloidal direction using a large device current.

In conclusion, we predict novel methods for reducing flux noise in superconducting devices: We demonstrated with explicit numerical calculations that enhancing ferromagnetic correlation between wire surfaces, 
or engineering a poloidal spin texture allows the reduction of the flux noise power due to electron spins by several orders of magnitude. The remaining noise power, due to lattice nuclear spins, can be as much as $20$ times smaller than the noise observed in the current state of the art devices.

\begin{acknowledgments}
  We wish to thank G. Aeppli, M.H. Amin, T. Lanting, M. Le Dall, and
  D. Stephen for useful discussions. This research was supported by
  the Natural Sciences and Engineering Research Council of Canada
  through its Discovery and Engage programs.
\end{acknowledgments}


\begin{thebibliography}{99}
   
\bibitem{clarke11} J. Clarke, {\it SQUIDs for Everything, an interview with Fabio Pulizzi}, Nat. Mater. {\bf 10}, 262 (2011).

\bibitem{everitt11} C. W. F. Everitt {\it et al.}, \prl {\bf 106}, 221101 (2011).

\bibitem{koch83} R.H. Koch, J. Clarke, W.M. Goubau, J.M. Martinis, C.M. Pegrum, and D. J. Van Harlingen, J. Low Temp. Phys. {\bf 51}, 207 (1983).

\bibitem{wellstood87} F.C. Wellstood, C. Urbina, and J. Clarke, \apl
  {\bf 50}, 772 (1987).

\bibitem{yoshihara06} F. Yoshihara, K. Harrabi, A.O. Niskanen, Y.
  Nakamura, and J.S. Tsai, \prl {\bf 97}, 167001 (2006).

\bibitem{bialczak07} R.C. Bialczak, R. McDermott, M. Ansmann, M.
  Hofheinz, N. Katz, E. Lucero, M. Neeley, A.D. O'Connell, H. Wang,
  A.N. Cleland, and J.M. Martinis, \prl {\bf 99}, 187006 (2007).

\bibitem{lanting09} T. Lanting {\it et al.}, \prb {\bf 79} 060509 (2009).

\bibitem{anton13} S.M. Anton, J.S. Birenbaum, S.R. O'Kelley, V.
  Bolkhovsky, D.A. Braje, G. Fitch, M. Neeley, G.C. Hilton, H.-M. Cho,
  K.D. Irwin, F.C. Wellstood, W.D. Oliver, A. Shnirman, and John
  Clarke, \prl {\bf 110}, 147002 (2013).

\bibitem{yoshihara10} F. Yoshihara, Y. Nakamura, and J.S.~Tsai, \prb {\bf 81}, 132502 (2010).

\bibitem{yan12} F. Yan, J. Bylander, S. Gustavsson, F. Yoshihara, K. Harrabi, D.G. Cory, T.P. Orlando, Y. Nakamura, J.-S. Tsai, and W.D. Oliver, \prb {\bf 85}, 174521 (2012).

\bibitem{clarke08} J. Clarke and F.K. Wilhelm, Nature {\bf 453}, 1031 (2008).

\bibitem{stern14} M. Stern, G. Catelani, Y. Kubo, C. Grezes, A. Bienfait, D. Vion, D. Esteve, and P. Bertet, \prl {\bf 113}, 123601 (2014).

\bibitem{slichter12}  D.H. Slichter, R. Vijay, S.J. Weber, S. Boutin, M. Boissonneault, J.M. Gambetta, A. Blais, and I. Siddiqi, \prl {\bf 109}, 153601 (2012).

\bibitem{sendelbach08} S. Sendelbach, D. Hover, A. Kittel, M. M\"{u}ck, J.M. Martinis, and R. McDermott, \prl {\bf 100}, 227006 (2008).

\bibitem{koch07} R.H. Koch, D.P. DiVincenzo, and J. Clarke, \prl {\bf 98}, 267003 (2007).

\bibitem{desousa07} R. de Sousa, \prb {\bf 76}, 245306 (2007).

\bibitem{faoro08} L. Faoro and L.B. Ioffe, \prl {\bf 100}, 227005 (2008). 

\bibitem{choi09} S.K. Choi, D.-H. Lee, S.G. Louie, and J. Clarke, \prl {\bf 103}, 197001 (2009).

\bibitem{chen10} Z. Chen and C.C. Yu, \prl {\bf 104}, 247204 (2010).

\bibitem{lee14} D. Lee, J.L. DuBois, and V. Lordi, \prl {\bf 112}, 017001 (2014).

\bibitem{sleator85} T. Sleator, E.L. Hahn, C. Hilbert, and J. Clarke, \prl {\bf 55}, 1742 (1985).

\bibitem{dube01} M. Dub\'{e} and P. Stamp, Chem. Phys. {\bf 268}, 257 (2001).

\bibitem{rose01} G. Rose and A.Y. Smirnov, J. Phys.: Condens. Matter {\bf 13}, 11027 (2001).

\bibitem{wu12} J. Wu and C.C. Yu, \prl {\bf 108}, 247001 (2012).

\bibitem{lanting14} T. Lanting, M.H. Amin, A.J. Berkley, C. Rich,
  S.-F. Chen, S. LaForest, and R. de Sousa, \prb {\bf 89}, 014503
  (2014).

\bibitem{zhang98} W. Zhang and D.G. Cory, \prl {\bf 80}, 1324 (1998). 

\bibitem{desousa09} R. de Sousa, Topics Appl. Physics {\bf 115}, 183 (2009). 

\bibitem{bylander11} J. Bylander, S. Gustavsson, F. Yan, F. Yoshihara, K. Harrabi, G. Fitch, D. G. Cory, Y. Nakamura, J.-S. Tsai, and W.D. Oliver, Nature Phys. {\bf 7}, 565 (2011).

\bibitem{atalaya14} J. Atalaya, J. Clarke, G. Sch\"{o}n, and A. Shnirman, \prb {\bf 90}, 014206 (2014).

\bibitem{de14} A. De, \prl {\bf 113}, 217002 (2014).

\bibitem{smithhisler96} C. Smithhisler, M. Kamon, L.M. Siveira and J. White, \href{http://www.layouteditor.net/wiki/FastHenry}{{\it FastHenry User's Guide Version 3.0}} (Research Laboratory of Electronics, Department of Electrical Engineering and Computer Science, Massachusetts Institute of Technology, 1996).

\bibitem{anton13b} S.M. Anton, I.A.B. Sognnaes, J.S. Birenbaum, S.R. O'Kelley, C.J. Fourie, and John Clarke, Supercond. Sci. Technol. 
{\bf 26}, 075022 (2013). 

\bibitem{chaikin00} Chapter 4 in P.M. Chaikin and T.C.
  Lubensky, {\it Principles of condensed matter physics} (Cambridge
  University Press, Cambridge U.K., 2000).

\bibitem{mermin66} N.D. Mermin and H. Wagner, \prl {\bf 17}, 1133 (1966).

\bibitem{jackson99} J.D. Jackson, {\it Classical Electrodynamics}, third edition (Wiley, New York, USA 1999).

\bibitem{fluxinductancenote} The flux-inductance theorem (proved in Section 5.17 of Ref.~\onlinecite{jackson99}) states that $\Phi_i={\cal H}_{{\rm mag}}/I_{{\rm SC}}$, where 
${\cal H}_{{\rm mag}}=+\bm{\mu}_i\cdot \bm{B}(\bm{R}_i)$ is the total energy required to produce the spin-dipole configuration (including the energy necessary to create the spin, see Section 5.16 of Ref.~\onlinecite{jackson99}). In Ref.~\onlinecite{lanting14} we missed a $-$ sign in the application of the flux-inductance theorem.

\bibitem{rhoderick62} E.H. Rhoderick and E.M. Wilson, Nature {\bf 194}, 1167 (1962).

\bibitem{vanduzer99} T. van Duzer and C.W. Turner, {\it Principles of
    superconductive devices and circuits} (Prentice Hall, New Jersey,
  USA 1999).

\bibitem{bouchiat09} V. Bouchiat, Supercond. Sci. Technol. {\bf 22}, 064002 (2009).

\bibitem{brown56} L.C. Brown {\it et al.}, J. Chem. Phys. {\bf 24}, 751 (1956).

\bibitem{sheriff51} R.E. Sheriff and D. Williams, Phys. Rev. {\bf 82}, 651 (1951).

\bibitem{baglin97} C.M. Baglin, Nuclear Data Sheets {\bf 80}, 1 (1997).

\bibitem{parkin91} S.S.P. Parkin, R. Bhadra, and K. P. Roche, \prl {\bf 66}, 2152 (1991).

\end{thebibliography}
\end{document}